\documentclass[draft]{agujournal2019}
\usepackage{url} %this package should fix any errors with URLs in refs.
\usepackage{lineno}
\usepackage[inline]{trackchanges} %for better track changes. finalnew option will compile document with changes incorporated.
\usepackage{soul}
\usepackage{natbib}

\journalname{JGR - Machine Learning and Computation}

\begin{document}

\title{Estimating carbon pools in the European Shelf sea environment: replacing reanalysis by model-informed machine learning?}
\authors{Jozef Sk\'akala \\ 
\emph{Plymouth Marine Laboratory, Plymouth, UK, \\
National Centre for Earth Observation, Plymouth, UK.}}

%\affiliation{1}{Plymouth Marine Laboratory, Plymouth, UK,}
%\affiliation{2}{National Centre for Earth Observation, Plymouth, UK.}

%
%  \note[editor]{The note}
%  \annote[editor]{Text to annotate}{The note}
%  \add[editor]{Text to add}
%  \remove[editor]{Text to remove}
%  \change[editor]{Text to remove}{Text to add}
%
% complete documentation is here: http://trackchanges.sourceforge.net/
%%%%%%%

\draftfalse

%% Enter journal name below.
%% Choose from this list of Journals:
%
% JGR: Atmospheres
% JGR: Biogeosciences
% JGR: Earth Surface
% JGR: Oceans
% JGR: Planets
% JGR: Solid Earth
% JGR: Space Physics
% Global Biogeochemical Cycles
% Geophysical Research Letters
% Paleoceanography and Paleoclimatology
% Radio Science
% Reviews of Geophysics
% Tectonics
% Space Weather
% Water Resources Research
% Geochemistry, Geophysics, Geosystems
% Journal of Advances in Modeling Earth Systems (JAMES)
% Earth's Future
% Earth and Space Science
% Geohealth
%
% ie, \journalname{Water Resources Research}

\correspondingauthor{Jozef Sk\'akala}{jos@pml.ac.uk}

%\begin{keypoints}
%\item enter point 1 here
%\item enter point 2 here
%\item enter point 3 here
%\end{keypoints}

%\begin{keypoints}
%\item Physics-biogeochemistry models can be used to train machine learning models to derive marine carbon pools from observations.
%\item Such machine learning model, when applied to reanalysis inputs, reproduces well the reanalysis for carbon pools.
%\item We propose to use machine learning as a computationally cheap alternative to reanalysis, wherever observations are absent or %uncertain.
%\end{keypoints}

%\linenumbers

\begin{keypoints}
%\item Physics-biogeochemistry models can be used to train machine learning (ML) models to derive marine carbon pools from observations.
%\item Such machine learning model, when applied to reanalysis inputs, reproduces well the reanalysis for carbon pools.
\item We present a marine-model-consistent, machine-learning-based reconstruction of carbon pools from observations.
\item By feeding reanalysis inputs into the ML model, its predictions reproduce several of the carbon pools from the same reanalysis. 
\item We propose using ML both as an efficient alternative to reanalysis and as a tool to simulate what-if scenarios.
\end{keypoints}

\begin{abstract}
%Shelf seas are important for carbon sequestration and carbon cycle, but available in situ, or satellite data for carbon pools in the shelf sea environment are often sparse, or highly uncertain. Alternative can be provided by reanalyses, but these are often expensive to run. We propose to use an ensemble of neural networks (NN) to learn from a coupled physics-biogeochemistry model the relationship between the directly observable variables and carbon pools. We demonstrate for North-West European Shelf (NWES) sea environment, that when the NN trained on a model free run simulation is applied to the NWES reanalysis, it is capable to reproduce the reanalysis outputs for carbon pools. Moreover, unlike the existing NWES reanalysis, the NN ensemble is also capable to provide uncertainty information for the pools. We focus on explainability of the results and demonstrate potential use of the NNs for future climate what-if scenarios. We suggest that model-informed machine learning presents a viable alternative to expensive reanalyses and could complement observational data, wherever they are missing and/or highly uncertain.
Shelf seas are important for the economy and the carbon cycle, but shelf sea observations for carbon pools are often sparse, or highly uncertain. An alternative can be provided by carbon reanalyses (whether assimilating proxy variables, such as chlorophyll-$a$, or directly carbon), but these are often expensive to run. We propose to use a computationally cheap ensemble of neural networks (i.e. deep ensemble) to learn the relationship between the directly observable (atmospheric, riverine and ocean) variables and marine carbon pools from a coupled physics-biogeochemistry model. The deep ensemble was trained on a North-West European Shelf (NWES) physical-biogeochemistry model free run simulation. 
After training, the deep ensemble was run using inputs from the NWES reanalysis instead of the free run, demonstrating that it can efficiently predict several NWES carbon pools (e.g., detritus, zooplankton, heterotrophic bacteria) in much better agreement with the reanalysis than the free run, while also providing uncertainty information. We further show that the deep ensemble performs similarly well when it is driven directly by the observations assimilated into the reanalysis, with the limitation that carbon pools can then be predicted only at the observed locations and times.
%After training, the deep ensemble was run using inputs from the NWES reanalysis instead of the free run, demonstrating that it is capable of efficiently predicting several NWES carbon pools (e.g., detritus, zooplankton, heterotrophic bacteria) in much better agreement with the reanalysis outputs than the free run, while additionally providing uncertainty information.
%After training, the deep ensemble was run using inputs from the NWES reanalysis instead of the free run, demonstrating that it is capable to efficiently predict several NWES carbon pools (e.g. detritus, zooplankton, heterotrophic bacteria) in much better agreement with the reanalysis outputs than the free run and additionally provide uncertainty information. 
%We also demonstrate that the deep ensemble performs similarly well when using directly the observations assimilated in the reanalysis as inputs, with the limitation that carbon pools are predicted only at the observed locations and times. 
We focus on explainability of the results and demonstrate potential use of the deep ensembles for future climate what-if scenarios. We suggest that model-informed machine learning presents a viable alternative to expensive reanalyses and could complement observations, wherever they are missing and/or highly uncertain.
\end{abstract}

\section*{Plain Language Summary}
%The ocean absorbs 30\% of atmospheric carbon emissions. Monitoring marine carbon pools is essential to understand how carbon cycles in the ocean, potentially being exported into the deep sea and buried in the sediment. Shelf seas, due to high concentrations of marine life, play a disproportionate role in carbon uptake and cycling. However shelf sea observations for different carbon pools can be often sparse, or uncertain, and marine ecosystem models can have substantial biases and uncertainties as well. Models and observations can be combined through data assimilation for their mutual benefit to provide reanalysis of shelf sea carbon pools, but this is computationally expensive. We propose here to derive a wide range of carbon pools (detritus, dissolved organic carbon, zooplankton, heterotrophic bacteria, dissolved inorganic carbon) from more directly observable variables using machine learning emulating a coupled shelf sea physics-biogeochemistry model. We demonstrate that such machine learning models provide a cost-effective alternative to the reanalysis wherever carbon observations are sparse, or highly uncertain. The low computational cost of running machine learning models has also other advantages, such as being able to easily provide uncertainty information for the carbon pools, or be used to simulate a wide range of future climate what-if scenarios.  

The ocean absorbs approximately 30\% of carbon emitted into the atmosphere. Monitoring marine carbon pools is essential to understand how carbon cycles within the ocean, including cross-shelf exchange between shelf seas and the open ocean, export to the deep sea and burial in sediments. Shelf seas, owing to land-sea exchange, the solubility pump, and high biological productivity, play a disproportionate role in carbon uptake and cycling. However, observations of different carbon pools in shelf seas are often sparse or uncertain, and marine ecosystem models can also exhibit substantial biases and uncertainties. Models and observations can be combined through data assimilation to produce reanalyses of shelf sea carbon pools, but this approach is computationally expensive. Here, we propose deriving a range of carbon pools (detritus, dissolved organic carbon, zooplankton, heterotrophic bacteria, dissolved inorganic carbon) from more directly observable variables using machine learning (ML) to reproduce their relationships from a coupled physics–\-biogeochemistry model. We demonstrate that such ML models provide a cost-effective alternative to reanalysis where carbon observations are sparse or highly uncertain. The low computational cost of using ML for inference offers additional advantages, including enabling straightforward uncertainty quantification and facilitating simulations of a wide range of future climate what-if scenarios.

%% ------------------------------------------------------------------------ %%
%  Title
%
% (A title should be specific, informative, and brief. Use
% abbreviations only if they are defined in the abstract. Titles that
% start with general keywords then specific terms are optimized in
% searches)
%
%% ------------------------------------------------------------------------ %%

% Example: \title{This is a test title}

\section{Introduction}

About 30\% of carbon emitted into the atmosphere ends up absorbed by the ocean, where it circulates in a multitude of organic and inorganic forms (\citet{friedlingstein2024global}). The inorganic carbon is assimilated by autotrophs during photosynthesis, which is followed by marine food web interactions distributing it between many other pools, such as higher trophic level species (e.g. zooplankton, heterotrophic bacteria, fish), non-living organic forms (e.g. detrital, dissolved organic carbon), or the dissolved and particulate inorganic carbon (e.g. \citet{emerson2008chemical}). Part of the marine carbon gets eventually deposited to the seafloor and buried, which helps to mitigate the anthropogenically driven climate change (\citet{volk1985ocean}). Understanding the ocean carbon cycle is therefore essential for better understanding Earth's climate response to atmospheric carbon emissions, both in past and future projections. Although the coastal oceans and shelf seas cover only $10-15\%$ of the global ocean, their impact on the global carbon cycle is disproportionately large (e.g. \citet{roobaert2019spatiotemporal, roobaert2024unraveling, cao2020diagnosis, chau2022seamless, dai2022carbon}). The North-West European Shelf (NWES), comprising a range of seas including the North Sea, Celtic Sea, Irish Sea, and the English Channel, is of major importance for the economy and the carbon cycle, as it is a highly biologically productive area with a significant impact on carbon sequestration and transport (\citet{borges2006carbon, jahnke2010global, legge2020carbon}).  
%Carbon that gets absorbed by the sea water in its inorganic form gets used by autotrophs during photosynthesis. This is followed by material flows distributing it between many other pools, i.e. higher trophic level species (e.g. zooplankton, heterotrophic bacteria, fish), non-living organic forms (e.g. detrital, dissolved organic carbon), or the dissolved and particulate inorganic carbon pools (e.g. \citet{emerson2008chemical}). Estimating these carbon pools on the NWES therefore provides an important insight into the NWES carbon cycle. 

A major source of information for the NWES are ocean surface observations derived from the satellite measurements. The biogeochemistry variable most typically derived from satellite ocean color is the surface chlorophyll-$a$ pigment concentrations (e.g. \citet{groom2019satellite, sathyendranath2019ocean}), which contain only indirect information about the NWES carbon pools. 
%Variety of methods how to estimate carbon from satellite more directly have been proposed for a range of pools \citep{brewin2021sensing}, including total particulate organic carbon (POC, e.g. \citet{evers2017validation, le2018color}), total phytoplankton carbon (e.g. \citet{roy2017size, sathyendranath2020reconciling}), dissolved organic carbon (DOC, e.g. \citep{matsuoka2017pan, laine2024machine}), and even indirect methods to estimate zooplankton carbon (e.g. \citet{stromberg2009estimation, behrenfeld2019global}) and of some bacterial species \citep{grimes2014viewing, racault2019environmental} have been proposed.
A variety of methods to estimate carbon from satellites more directly have been proposed for a range of pools (\citet{brewin2021sensing}), including total particulate organic carbon (POC; e.g., \citet{evers2017validation, le2018color}), total phytoplankton carbon (e.g., \citet{roy2017size, sathyendranath2020reconciling}), dissolved organic carbon (DOC; e.g., \citet{matsuoka2017pan, laine2024machine}), as well as indirect approaches to estimate zooplankton carbon (e.g., \citet{stromberg2009estimation, behrenfeld2019global}) and carbon associated with some bacterial species (\citet{grimes2014viewing, racault2019environmental}).
However these satellite algorithms have been most commonly developed for the global ocean, representing mostly open ocean conditions. Some shelf sea and coastal products exist for specific pools (e.g. DOC, \citet{mannino2008algorithm, matsuoka2017pan}), but have been developed for areas far from the NWES. Thus, apart of the relatively high uncertainty associated with many of those products, they might not be particularly suitable for the NWES.
Beyond these satellite products the only available observations are in situ measured data. These cover quite well specific variables related to dissolved inorganic carbon (DIC), such as CO$_{2}$ fugacity and partial pressure (\citet{bakker2014update}), which are typically governed by well-established thermodynamic relationships and can be either directly measured or robustly derived from direct observations. However organic carbon pools are typically operationally defined, meaning their quantification depends on methodological choices (e.g., filtration thresholds, oxidation techniques, or analytical protocols). This leads to greater heterogeneity and sparsity across datasets, and those data become too limited to provide us with a more detailed understanding of organic NWES carbon pools (for more general overview of in situ observational capacity in marine biogeochemistry see e.g. \citet{telszewski2018biogeochemical}).

A more comprehensive representation of the system is provided by marine biogeochemistry models (\citet{wakelin2012modeling}). However, in the NWES these models can exhibit substantial biases and uncertainties. For example, the European Regional Seas Ecosystem Model (ERSEM; \citet{baretta1995european, butenschon2016ersem}), which is used operationally in the region, shows pronounced seasonal errors in phytoplankton phenology. Specifically, it simulates overly intense and delayed spring blooms, alongside near-vanishing phytoplankton concentrations during winter (see \citet{skakala2022impact, skakala2024how} for discussion). Because phytoplankton form the base of the marine food web, such biases propagate through the ecosystem and influence carbon cycling (\citet{skakala2022impact, skakala2024how}). An intermediate approach between sparse observations and imperfect free-running models (unconstrained by the observations through data assimilation) is provided by the NWES Copernicus reanalysis (\citet{kay2016north, kay2019north}). This framework constrains the model primarily with satellite observations while delivering a range of outputs for carbon pools.
In the NWES reanalysis, ERSEM biogeochemistry is constrained through the assimilation of satellite phytoplankton functional type (PFT) chlorophyll-$a$ (\citet{brewin2017uncertainty, skakala2018assimilation}). This substantially reduces key biases in phytoplankton phenology (\citet{skakala2018assimilation, kay2016north}), making the reanalysis outputs generally more reliable than those from the free-running model. Given the scarcity of robust and comprehensive observations, we treat the reanalysis carbon pool estimates in this study as the dataset closest to “ground truth”. Nevertheless, generating such reanalyses is computationally expensive, particularly for complex biogeochemical models. To maintain computational feasibility, several simplifications are implemented. Most notably, only biogeochemical variables directly related to the assimilated PFT chlorophyll-$a$ product, i.e. phytoplankton biomass, are directly constrained. Other variables (e.g., non-phytoplankton carbon pools) are impacted only through the model dynamical adjustment (\citet{skakala2018assimilation}).

%In this paper we propose an alternative: in line with the spirit of satellite retrieval algorithms we use a machine learning (ML) model to learn the relationship between the more directly observable variables from the satellite (such as sea surface temperature, ocean colour-derived PFT chlorophyll-$a$) and the unobserved carbon pools (for examples of ML-based satellite algorithms for marine carbon pools see \citet{sauzede2020estimation, zemskova2022deep, li2024advanced, laine2024machine, zhang2025review}), but in our case the relationship is learned directly from the underlying NWES coupled physics-biogeochemistry model. 
In this paper, we propose an alternative approach. In the spirit of satellite retrieval algorithms, we use a machine learning (ML) model to learn the relationship between directly observable satellite variables (such as sea surface temperature and ocean-colour-derived PFT chlorophyll-$a$) and unobserved carbon pools. (For examples of ML models estimating carbon pools see \citet{sauzede2020estimation, zemskova2022deep, li2024advanced, laine2024machine, zhang2025review}.) However, in our case, the relationship is learned directly from the underlying NWES coupled physics–biogeochemistry model.
In the coupled model such relationship emerges from a plethora of simulated processes, ranging from hydrodynamics to complex ecosystem interactions. Using the coupled model to inform the machine learning has several advantages: (i) there is no shortage of training data, i.e. the map from the observable variables to the carbon pools can be learned from the existing free run simulations, providing gap-free and abundant outputs, (ii) the coupled model is specifically calibrated for the NWES domain, and (iii) any carbon pool outputted by the model can be in theory predicted, including vertical distributions. 

The disadvantage of this approach is the assumption that the model-simulated relationship between the established satellite-observable variables (together with a range of model forcing data) and the derived carbon pools is sufficiently realistic. This means that any biases in the carbon pools modelled in the free run will be primarily due to model misrepresenting the dynamics of the observable variables (e.g. showing large seasonal biases in chlorophyll-$a$), rather than in the modelled relationship between those variables and the carbon pools. This is a strong assumption, however, it can be argued that similarly strong assumptions are effectively used in the NWES reanalysis itself, when many of the carbon pools are left to dynamically respond within the same coupled model to the assimilation updates to PFT biomass components and physical variables. On the other hand, data assimilation provides information about the uncertainty in the estimated carbon pools. This uncertainty comes from errors in the observational data and from different sources of uncertainty in the coupled physics–biogeochemistry model. These include uncertainty in model forcing, initial and boundary conditions, model structure, and model parameter values. We assume that the last two sources (model structure and parameter values) are major contributors to uncertainty in how the model maps observable variables to carbon pools. Our ML model learned only one version of this map (e.g. related to a specific set of marine model parameter values), and it cannot directly represent this uncertainty. We therefore conducted additional ensemble 1D simulations as an initial step toward quantifying the magnitude of this uncertainty, as described in the Results section.

%We thus performed additional ensemble 1D simulations to do some first steps towards estimating the scale of this uncertainty, as described in the Results section.

%On the other hand data assimilation represents the uncertainty of the estimated pools due to observational data uncertainty and a range of model uncertainties, originating from model forcing, initial and boundary conditions, model formulation and the uncertain model parameter values. We assume that the last two sources of uncertainty introduce uncertainty into the map between observable variables and carbon pools, learned from the model. We did in this work some initial steps towards estimating the scale of this uncertainty. 

In this paper, we demonstrate that a reasonably simple feed-forward neural network (NN) can successfully learn from the operational physics–biogeochemistry model for the NWES how to derive a range of surface carbon pools from several observable variables, forcing data, and structural variables such as latitude and longitude. The carbon pools are derived at a relatively coarser spatial (35 km) and temporal (10-day) resolution than that of the operational model outputs (7 km and daily), although this coarser resolution is entirely sufficient for regional climate studies. 
%When using inputs provided by the reanalysis rather than the model free run on which it was trained, the NN reproduces the same reanalysis for several ocean surface carbon pools reasonably well (i.e., much better than the free run).
When driven by inputs from the reanalysis rather than from the free-running model on which it was trained, the NN reproduces the reanalysis estimates of several surface ocean carbon pools reasonably well (i.e. substantially better than the free run). The same holds when the NN uses as inputs the observations directly assimilated into the reanalysis, as intended in its application. We provide insights into the explainability of these results and discuss interesting applications, such as running future climate what-if scenarios using lightweight ML models.

\section{Methodology}

\subsection{The physics-biogeochemistry coupled model}

The model used in NWES Copernicus reanalysis (\citet{kay2019north, kay2016north}) that we are referring to in this study is the physical model Nucleus for European Modelling of the Ocean (NEMO, \citet{madec2015nemo}), coupled to ERSEM (\citet{baretta1995european, butenschon2016ersem}) through Framework for Aquatic Biogeochemical Models (FABM, \citet{bruggeman2014general}). The NEMO ocean physics component is a finite difference, hydrostatic, primitive equation ocean general circulation model, here used with a 7km (AMM7) NWES configuration using the terrain following $z*-\sigma$ coordinates with 51 vertical layers (\citet{o2017co5, siddorn2013analytical}). ERSEM is a high complexity ecosystem model representing elemental cycles of carbon, nitrogen, phosphorus and silicon using variable stoichiometry. It represents four phytoplankton functional types (PFTs) which are mostly size-based (diatoms, nanophytoplankton, microphytoplankton and picophytoplankton), zooplankton in three functional types (mesozooplankton, microzooplankton and heterotrophic nanoflaggelates) and as a decomposer it includes heterotrophic bacteria (\citet{butenschon2016ersem}). The non-living organic matter is represented in three detrital forms (large, medium-size, small) and three forms of dissolved organic matter (labile, semi-labile and semi-refractory) (\citet{butenschon2016ersem}). 
%ERSEM uses variable stoichiometry representing biomass in carbon, nitrogen, phosphorus, for phytoplankton also chlorophyll and in case of diatoms also silicon. 
ERSEM uses variable stoichiometry to represent biomass in terms of carbon, nitrogen, and phosphorus, and for phytoplankton also chlorophyll-$a$, with silicon additionally included for diatoms.
ERSEM includes carbonate system as per \citet{artioli2012carbonate}, representing dissolved inorganic carbon (DIC) and total alkalinity as two independent state variables (from which one can derive pH and pCO$_{2}$ as diagnostic variables). 

%- discuss NEMO

%- discuss ERSEM

%- discuss the training data used

\subsection{The machine learning model}

\begin{figure}
\hspace{0.cm}
\noindent\includegraphics[width=13cm, height=9cm]{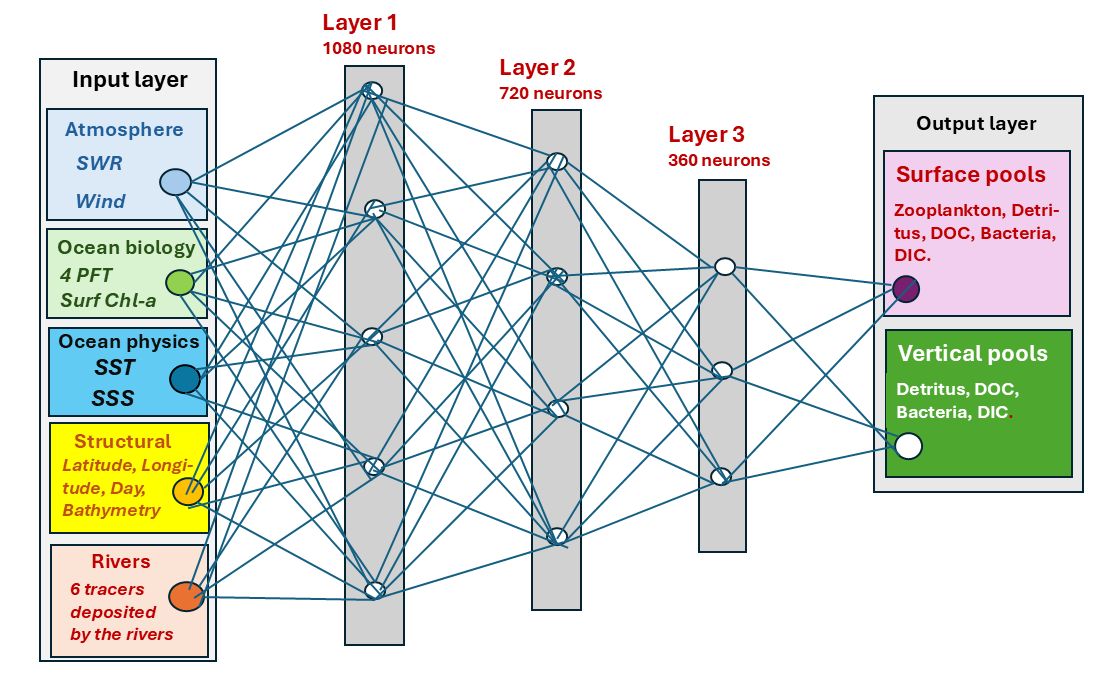}
\makeatletter 
\makeatother
\caption{A schematic representation of the feedforward NN model, its inputs and outputs. For simplicity the schematic representation merges the same types of inputs and outputs into one node on the diagram (i.e., in reality 18 inputs and 9 outputs were used rather than 5 input and 2 output nodes displayed in the schematic Figure). The abbreviations used are, SST: sea surface temperature, SSS: sea surface salinity, SWR: short-wave radiation, DOC: dissolved organic carbon, DIC: dissolved inorganic carbon.
%The 2016-2020 time-series of domain-averaged values of the training-validation data-set from the free run (red) and the reanalysis (blue). The different panels compare the biogeochemical variables of key interest: (i) the surface PFT chlorophyll directly corrected by the assimilation in the reanalysis and (ii) the NN-predicted surface carbon pools outputted by both the free run and the reanalysis.

}
\label{Fig.0.1}
\end{figure}

We employed a fully-connected feed-forward neural network (NN) implemented using TensorFlow Keras, consisting of three hidden layers with a total of 2160 neurons, arranged in a decreasing configuration of 1080, 720, and 360 neurons per layer (see Fig.\ref{Fig.0.1}). The NN model used Adam optimizer, root mean squared cost function and random initialization of weights based on normal distribution. To reduce overfitting and optimize performance we have used dropout function with 30\% of neurons randomly switched off at each learning step. This architecture was chosen after extensive testing of NN performance across range of different architectures (e.g. numbers of hidden layers) and hyperparameter values. %The 2016-2020 data-set was split into the 2016-2019 period used as the training data-set and 2020 year for validation. We have used a 15-member ensemble of the NN model realizations (i.e. deep ensemble) to boost the performance and introduce some estimate of (mostly epistemic) uncertainty. The ensemble members were naturally distinguished through the random model parameters, such as weight initialization, dropout function and other random procedures, such as splitting the training data into batches (batch size of 32 was used). Prediction was taken from the ensemble mean and uncertainty was evaluated as the ensemble standard deviation. 

We have used a 15-member ensemble of the NN model realizations (i.e. deep ensemble) to boost the performance and introduce some estimate of (mostly epistemic) uncertainty. The ensemble members were naturally distinguished through the random model parameters, such as weight initialization, dropout function and other random procedures, such as splitting the training data into batches (batch size of 32 was used). The ensemble members used for training the same number of epochs. The coefficient of variation of the final training loss across ensemble members was 5.8\%, indicating moderate optimization variability. Prediction was taken from the ensemble mean and uncertainty was evaluated as the ensemble standard deviation. 

The deep ensemble used as input features (i) observable ocean variables, i.e. sea surface PFT chlorophyll, sea surface temperature (SST) and sea surface salinity, (ii) ``structural'' inputs, i.e. latitude and longitude, bathymetry and the day of the year, (iii) atmospheric inputs, i.e. incoming short-wave radiation, surface wind speed, and (iv) riverine discharge data variables such as freshwater runoff, nutrients, oxygen, DIC (see also schematic representation in Fig.\ref{Fig.0.1}). The riverine inputs were considered non-zero in the 50 km neighborhood of the river mouths, as described in \citet{banerjee2025improved}. The deep ensemble predicted a number of carbon pools in the output layer, i.e. for (i) total surface zooplankton, (ii) total surface detritus, (iii) total surface dissolved organic matter, (iv) surface heterotrophic bacteria, (v) surface dissolved inorganic matter, and vertically averaged values for all these pools except for zooplankton. All input features and predicted outputs were standardized using z-score normalization $\hat{x}_{i} = (x_{i}-\mu_{x})/\sigma_{x}$, where $\mu_{x}$ and $\sigma_{x}$ denote the mean and standard deviation computed over the training dataset. This standardization mitigates scale differences between variables and prevents features or outputs with larger numerical ranges from disproportionately influencing the optimization process.

The training and validation (training-validation) data were: (i) for the ocean variables provided by the 5 year 2016-2020 NEMO-FABM-ERSEM free simulation (the time series of the domain averaged data are shown in Fig.\ref{Fig.0.5}), (ii) for the atmospheric variables by the corresponding 2016-2020 ERA-5 atmospheric product (https://www.ecmwf.int/\-en/forecasts/datasets/reanalysis-datasets/era5), forcing the oceanic model simulation, and (ii) for the river data by the riverine data-set originating from \citet{lenhart2010predicting}, also used to force the oceanic model simulation. The 2016-2020 data-set was split into the 2016-2019 period used as the training data-set and 2020 year for validation. 
%This selection of the training-validation split was motivated by separating the two data-sets in time, to maximize their independence by minimizing the impact of temporal auto-correlations. 
The training–validation split was chosen to separate the datasets temporally, thereby maximizing their independence and reducing the influence of temporal autocorrelation.

The NEMO-FABM-ERSEM configuration providing the simulations for the training-validation data has been described in \citet{skakala2019improved}. Similarly to the Copernicus reanalysis the model horizontal resolution was 7km with 51 vertical grid layers using a terrain-following z$^{*}$-$\sigma$ coordinate system. The model was forced by atmospheric data from the ERA-5 reanalysis and used lateral boundary conditions from the GloSea5 Seasonal Forecasting System (\citet{maclachlan2024global}); and from a reanalysis produced by the Danish Meteorological Institute for the Copernicus Marine Service. The simulation used NEMO model version 3.6, a development of the CO5 configuration explained in detail by O'Dea et al. (2017). The light in the ERSEM model was forced by a spectral bio-optical model described in \citet{skakala2019improved}, and both NEMO and ERSEM were forced by an updated version of the river discharge dataset from \citet{lenhart2010predicting}.  The simulation delivering the training-validation data-set has been performed in \citet{higgs2023ecosystem}, providing daily and 7km resolution outputs for all the relevant carbon pools. 
The training-validation data have been produced by coarsening the simulation outputs (through averaging) to a 10-day temporal scale and 35km (5x5 model pixel) spatial scale. The deep ensemble inference is consequently applied only to such coarser spatial and temporal scales, which are however considered sufficient to monitor carbon pools on the NWES. These coarser scales were chosen to reduce the data redundancy/duplication due to their spatial and temporal correlations, improving the efficiency of the training and validation of the ML model. The overall number of data-points in the coarsened training-validation data was over 500 000.

The performance of the relatively lightweight feed-forward neural network (with three hidden layers and $\sim$ 2200 neurons) could be improved by exploiting the spatial and temporal structure of the inputs, for example by using more sophisticated architectures such as convolutional neural networks (CNNs; \citet{li2021survey}), graph neural networks (GNNs; \citet{scarselli2008the}), or long short-term memory networks (LSTMs; \citet{hochreiter1997long, yu2019review}). However, to predict carbon pools directly from observations, these architectures would need to handle substantial time-varying gaps caused by cloud cover, atmospheric disturbances and low solar zenith angle, including seasonal gaps and the near-absence of data over large parts of the NWES in winter. Such seasonally biased data gaps — potentially correlated with the target ocean variables — may pose significant challenges for these architectures, especially if the training data-set is reasonably limited (e.g., \citet{zhou2020graph}). Another challenge is how to generalize models trained on gap-free free-run simulations (both spatially and temporally) so that they can make accurate predictions when using satellite data as input, which frequently contain substantial gaps. Finally, models that rely on spatial and temporal relationships (such as CNNs or GNNs) and are trained on the model free run may underperform when applied to reanalysis inputs because the spatial and temporal consistency of reanalysis variables is degraded by the intermittent availability of assimilated observations, leading to highly uneven impacts across the model domain. For these reasons, at this initial proof-of-concept stage, we chose to use a simpler perceptron neural network that predicts carbon pools using only inputs from the same location as the target value. In future work, however, it will be desirable to explore more complex architectures and/or explore alternative very recently developed physics-informed approaches (\citet{raissi2019a}) to predict target data from incomplete observations, such as 4DVarNet (\citet{fablet2021joint}).

%Since this could prove challenging for CNNs, or GNNs, the simpler feed-forward NN was preferable as an initial proof of concept in this study, as it focuses on local predictions wherever inputs are available ignoring t.

\subsection{The multi-decadal reanalysis}

\begin{figure}
\hspace{-1.8cm}
\noindent\includegraphics[width=17cm, height=17cm]{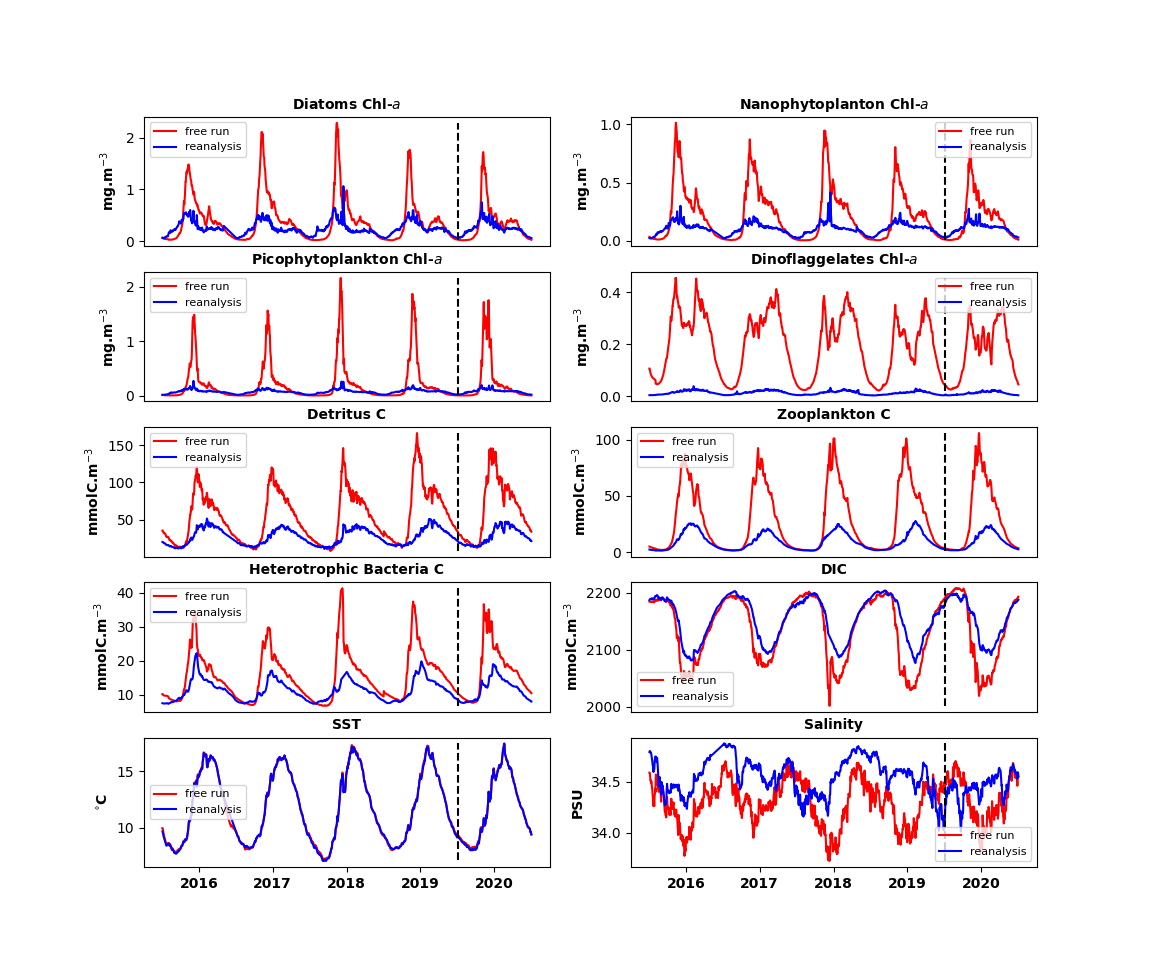}
\makeatletter 
\makeatother
\caption{
%The 2016-2020 time-series of domain-averaged values of the training-validation data-set from the free run (red) and the reanalysis (blue). The different panels compare the biogeochemical variables of key interest: (i) the surface PFT chlorophyll directly corrected by the assimilation in the reanalysis and (ii) the NN-predicted surface carbon pools outputted by both the free run and the reanalysis.
Time series (2016–2020) of domain-averaged values from the training and validation dataset derived from the free run (red) and the test data from the reanalysis (blue). The panels compare the key biogeochemical and physics variables of interest: (i) surface PFT chlorophyll, directly corrected by data assimilation in the reanalysis (first two rows), (ii) those (predicted by the NN) surface carbon pools which were outputted by both the free run and the reanalysis (rows three and four) and (iii) physics variables (SST and surface salinity), both assimilated in the reanalysis (bottom row). The vertical dashed lines show the separation between the training and validation data from the free run.
}
\label{Fig.0.5}
\end{figure}

The capability of the ML model to predict carbon pools using observational data as inputs was subsequently tested on a data-set provided by Copernicus multi-decadal reanalysis (\citet{kay2019north, kay2016north}). The reanalysis is based on a very similar NEMO-FABM-ERSEM model configuration as the free run used for training-validation. The NEMO-FABM-ERSEM model was constrained in the reanalysis through daily assimilation of SST from European Space Agency (ESA) Climate Change Initiative (CCI) v1.1 product, in situ SST from International Comprehensive Ocean-Atmosphere Data Set (ICOADS), temperature and salinity profiles from EN4
data (\citet{good2013en4}), and PFT (log-)chlorophyll from ESA CCI v3.1 data (\citet{sathyendranath2019ocean}). The data were assimilated using NEMOVAR (\citet{mogensen2012nemovar}), based on 3DVar approach similar to the one described in \citet{king2018improving, skakala2018assimilation}. Within ERSEM the assimilation of PFT chlorophyll-$a$ updates only the PFT biomass (all components chlorophyll-$a$, carbon, nitrogen, phosphorus, and for diatoms silicon), based on the forecast stoichiometry. The remaining ERSEM variables are not directly constrained by the assimilation and evolve only during the subsequent model simulation through dynamical adjustment. Fig.\ref{Fig.0.5} shows that the assimilated PFT chlorophyll-$a$, as well as the simulated surface carbon pools, differ substantially between the reanalysis and the free run that provided the training and validation data for the NN. This discrepancy is well documented in the literature (e.g., \citet{skakala2018assimilation}). Consequently, the reanalysis data constitute a highly non-trivial test of NN performance.

%The remaining ERSEM variables are unconstrained by the assimilation, and their values change only during the subsequent model simulation through the dynamical adjustment. 
%Fig.\ref{Fig.0.5} demonstrates that the assimilated PFT chlorophyll-$a$, as well as the simulated surface carbon pools, show major differences between the reanalysis and the free run, which provided the training and validation data for the NN (this is already well known from the literature, e.g. see \citet{skakala2018assimilation}). This means the reanalysis data provide highly non-trivial test of the NN performance. 
%The Fig.A1 of the Appendix demonstrates that the assimilated biogeochemistry variables, as well as the simulated surface carbon pools show major differences between the reanalysis and the free run, which provided the training and validation data for the NN.

The reanalysis carbon can be validated with independent data, wherever they are available. For the relatively abundant ICES pCO2 data-set this was already done in \citet{kay2016north}, demonstrating a good skill. We have done here an additional comparison of the reanalysis outputs with OC-CCI v4.2 satellite product for the total POC (\citet{stramski2008relationships, evers2017validation}), comprising aggregate carbon of phytoplankton, zooplankton, detritus and bacteria pools. The results are shown in Fig.A1 of the Appendix. They indicate that the POC in the reanalysis has significant negative biases relative to the satellite product, but the way the POC values are distributed across the domain is to a degree similar between both products.
%distributions are reasonably comparable between the reanalysis and satellite-derived product with some overall negative biases of the reanalysis. 
Satellite-derived DOC from \citet{laine2024machine} was also available, unfortunately DOC was not outputted by the reanalysis and could not be compared to the satellite observations. However issues (major positive biases in hundreds of percents of the observed values) with ERSEM DOC have been found (\citet{neccton2025})
and have been also confirmed here (not shown). Although DOC pool is included here to demonstrate the ML capability to learn it from the model, it is recognized that as an end-product it would likely have at this stage only limited value.

%, and (ML-derived) product for dissolved organic carbon (DOC) \cite{}, both covering also the NWES domain. The reanalysis does provide the equivalent of the POC pool, and the POC validation is shown in Fig.\ref{}, demonstrating some overall negative biases, but generally reasonable comparison of the spatial POC distributions between the reanalysis and satellite product. Unfortunately DOC was not outputted by the reanalysis and could not be compared to the satellite observations, however issues (major positive biases in hundreds of percents of the observed values) with ERSEM DOC are known \cite{} and have been confirmed here (not shown). Although DOC pool is included here to demonstrate the ML capability to learn it from the model, it is recognized that as an end-product it would likely have at this stage only limited value.

%- discuss the DA system

%- the reanalysis validation

\subsection{Experiments and validation metrics\label{sec-exp}}

The model performance on the independent data produced by the reanalysis was evaluated by comparing the spatial distributions of 2016-2020 time-averaged reanalysis and predicted carbon pools, indicating the overall biases of the prediction. A separate metrics of Bias-Corrected Root-Mean Square Difference (BC-RMSD) was applied, defined as 
\begin{equation}\label{eq1}
\hbox{BC-RMSD} = \sqrt{\langle (X_{1} - X_{2} - (\langle X_{1}\rangle - \langle X_{2} \rangle))^{2}\rangle}, 
\end{equation}
where $X_{1}, X_{2}$ are the two compared data-sets. The BC-RMSD \% improvement generated by data 1 (with $\hbox{BC-RMSD}_{1}$) relative to data 2 (with $\hbox{BC-RMSD}_{2}$) is defined as
\begin{equation}\label{eq2}
\hbox{BC-RMSD}_{imp} = 100\cdot\frac{\hbox{BC-RMSD}_{1} - \hbox{BC-RMSD}_{2}}{\hbox{BC-RMSD}_{2}}.  
\end{equation}

To demonstrate the ultimate goal of this work — namely, predicting carbon pools directly from observations — we conducted an additional experiment in which the SST and PFT chlorophyll-$a$ inputs from the reanalysis were replaced with the corresponding satellite observations assimilated into the reanalysis. Because the satellite products contain substantial data gaps, it was necessary to define a minimum data-availability threshold for aggregating the observations to the 35 km spatial and 10-day temporal resolution required by the NN. For PFT chlorophyll-$a$, the minimum threshold was set to 10\% of the maximum possible number of OC-CCI observations within a 35 km, 10-day grid cell, given the native $\sim$ 5 km daily resolution of the OC-CCI product. For SST, the threshold was reduced to 5\% in order to retain a sufficient number of samples. This lower threshold is justified because the assimilated merged CCI v1 SST product already combines data from multiple satellites, thereby reducing observational uncertainty. After applying these processing steps, just over 50,000 observation-based samples remained for NN testing over the 2016–2020 period — approximately an order of magnitude fewer than in the reanalysis-based test dataset.

%To demonstrate the ultimate goal of this work, which is predicting the carbon pools directly from the observations, we have replaced in another experiment the SST and PFT chlorophyll-$a$ inputs from the reanalysis with the satellite observations assimilated into the reanalysis. As the satellite data have many gaps, one has to make the choice about the minimum number of data-points to provide satellite estimate for the 35 km and 10-day resolution of the inputs used by the NN. For PFT chlorophyll-$a$ the minimum number of data-points was 10\% relative to what would be the maximum number of OC-CCI data fitting into a 35km and 10-daily pixel, given the $\sim$ 5km and daily resolution of the OC-CCI product. For SST we lowered the threshold to 5\% (to maintain sufficient number of data), which can be justified by the fact that in the assimilated merged CCI v.1 SST product already blends data from multiple satellites, reducing their uncertainties. After these processing steps there was left over 50 000 observational-based test data for the NN for the 2016-2020 period, about a factor 10 less than in the test data based on the reanalysis. 

We have also used the deep ensemble to predict hypothetical what-if scenarios. Two scenarios were chosen for the ensemble prediction: (i) One scenario, where PFT chlorophyll is scaled down from the 2016-2020 reanalysis value gradually to zero, maintaining the same PFT community structure and spatio-temporal distributions as in the reanalysis. This meant taking $\gamma.\hbox{Chl}(x,t)$, with $\hbox{Chl}(x,t)$ being the reanalysis chlorophyll-$a$ and $\gamma$ a scaling parameter lowered from 1 to 0. (ii) Another scenario, where the ratio of large phytoplankton (sum of diatoms and dinoflaggelates) to total phytoplankton chlorophyll was gradually scaled from the 2016-2020 reanalysis down to zero, but maintaining the same total chlorophyll concentration and spatio-temporal distributions as in the 2016-2020 reanalysis. This means we used $\gamma_{1}.\hbox{Chl}_{1}(x,t)$ and $\gamma_2.\hbox{Chl}_{2}(x,t)$, where $\hbox{Chl}_{1}(x,t)$ is the large phytoplankton chlorophyll-$a$ (diatoms and dinoflaggelates) and $\hbox{Chl}_{2}(x,t)$ is the smaller phytoplankton chlorophyll-$a$ (nanophytoplankton and picophytoplankton). The $\gamma_{1}$ was a scaling parameter gradually reduced from 1 to 0 and $\gamma_{2}$ another parameter proportionally increased above 1 to maintain that 
\[\gamma_{1}.\hbox{Chl}_{1}(x,t)+\gamma_{2}.\hbox{Chl}_{2}(x,t) = \hbox{Chl}_{1}(x,t) + \hbox{Chl}_{2}(x,t) = \hbox{Chl}(x,t).\] 
In both cases the NN model inputs other than PFT chlorophyll were kept the same as in the reanalysis. These scenarios (defined by the NN inputs) were motivated by certain aspects of future climate projections for the NWES (\citet{wakelin2015modelling}), but are obviously major simplifications, e.g. there is no guarantee that the scenarios are sufficiently self-consistent, as they did not come from a model simulation.

%designed to represent certain aspects of future climate where net primary production is substantially reduced and phytoplankton community experiences shift to small species, both as a result of surface nutrient depletion \cite{}.  

%- the design and reasoning behind it

%- splitting the training/validation data and tests

%- uncertainty and ensemble methods

%- experiments and metrics used

\section{Results and discussion}

The deep ensemble performance on validation data, as shown in Tab.\ref{Tab.1}, is very good, with $R^{2}$ of the surface pools between 0.83 and 0.89, and for the vertically averaged pools in the 0.9-0.92 range, except for the vertically averaged DIC, where it was lower ($R^{2}=0.68$).

\begin{table}[h]

%\centering

\caption{
Ensemble-averaged skill of the deep ensemble on the 2020 free-run validation data, measured for each variable using the R$^{2}$ score, the RMSD, and the percentage RMSD relative to the natural variability of that variable, as quantified by the standard deviation ($\sigma$) of the validation data.}
%Ensemble-averaged skill of the deep ensemble on the 2020 free run validation data, as measured for each variable by the R$^{2}$ score, the RMSD, and the percentage RMSD when compared to the the natural variability of that variable as measured by the standard deviation ($\sigma$) of the validation data.}
\hspace{-1cm}
\label{Tab.1}
\footnotesize
\begin{tabular}{l | c c c c c | c c r}
\hline
 ~ & \multicolumn{5}{c|}{{\bf Surface C}} & \multicolumn{3}{c}{\bf {Vertically averaged C} ~~}  \\
\hline

{\bf metric}  & {\bf Detritus} & {\bf DOC} & {\bf Zooplankton} & {\bf Het. bacteria}& {\bf DIC} & {\bf DOC} & {\bf Het. bacteria} & {\bf DIC} \\ 
\hline
R$^{2}$ & 0.83 & 0.87 & 0.86 & 0.85 & 0.89 & 0.92 & 0.9 & 0.68 \\
\hline
RMSE (mmolC/m$^{3}$) & 28.2 & 252.6 & 12.8 & 4.3 & 24.4 & 124.5 & 2.2 & 94.2 \\
\hline
RMSE $\sigma$-norm. (\%) & 41.5 & 36 & 37.1 & 38.3 & 33.4 & 28.6 & 31.6 & 56.1 \\
\hline
\end{tabular}
\end{table}

%\documentclass{article}
% optional for page layout

\begin{figure}
\hspace{-2.5cm}
\noindent\includegraphics[width=19cm, height=13cm]{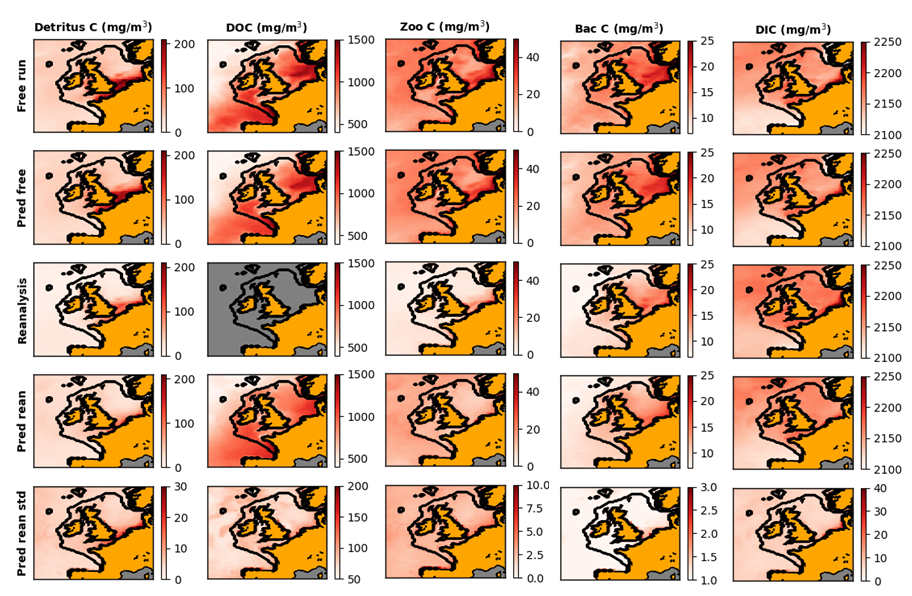}
\caption{The 2016-2020 average values of different estimated carbon pools comprising ocean surface concentrations of detritus (first column), DOC (second column), zooplankton (third column), heterotrophic bacteria (fourth column) and DIC (fifth column). Compared are the NEMO-FABM-ERSEM free run providing the training and validation data (first row), the prediction of these data by the mean of the deep ensemble (second row), the Copernicus reanalysis concentrations from \citet{kay2016north} (third row) and the analogue of these reanalysis concentrations predicted by the mean of the deep ensemble from the reanalysis inputs (fourth row). In the bottom, fifth row the panels show the uncertainties of the estimated concentrations from the fourth row obtained as the standard deviation of the deep ensemble averaged in time. The reanalysis DOC is masked, since it was not available in the reanalysis outputs.}
\label{Fig.1}
\end{figure}

The Fig.\ref{Fig.1} shows the performance of the deep ensemble on both, 2016-2020 free run training-validation data (comparing first and second rows), and the reanalysis test data (comparing third and fourth rows). The Figure compares the deep ensemble biases, as well as biases among the free run (first row) and the reanalysis (third row). It is demonstrated that except for DIC (fifth column) the surface concentrations of the carbon pools in the reanalysis are substantially lower than in the free run (see also Fig.\ref{Fig.0.5}). This is due to a reduction in phytoplankton concentrations caused by the assimilation of satellite data (see \citet{skakala2018assimilation, skakala2019improved}), which propagates to the other organic carbon pools. The deep ensemble, trained on the free-run data, is capable of capturing this pattern, as can be seen by comparing rows three and four in Fig.\ref{Fig.1}. In fact, for detrital matter the deep ensemble predicts a larger reduction than the reanalysis, whereas for zooplankton and bacteria the reduction is slightly smaller. However, in all cases the deep ensemble predictions are much closer to the reanalysis than the free run. Fig.\ref{Fig.1} also presents, in its bottom row, the uncertainty estimates obtained from the deep ensemble. It should be noted that these uncertainties reflect primarily the effects of neural-network optimization (and are therefore predominantly epistemic). They do not account for several important sources of uncertainty in the predicted carbon pools, such as those arising from the input variables or from the learned relationships between the observable variables and the carbon pools. 

%This is due to reduction in phytoplankton concentration caused by the assimilation of satellite data (see \citet{skakala2018assimilation, skakala2019improved}), that propagates to the other organic carbon pools. The deep ensemble, trained on the free run data, is capable to pick this pattern, as can be seen by comparing the rows three and four in Fig.\ref{Fig.1}. In fact for detrital matter the deep ensemble predicts larger reduction in detritus than the reanalysis, whilst for zooplankton and bacteria the reduction is slightly lower, but the deep ensemble predicted values are in all cases much closer to the reanalysis than the free run. 

\begin{figure}
\hspace{-2.5cm}
\noindent\includegraphics[width=18cm, height=11cm]{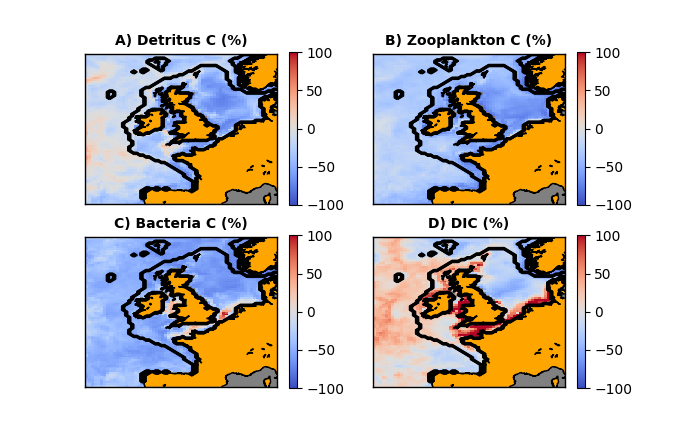}
\caption{The relative percentage (\%) improvement at ocean surface captured by BC-RMSD$_{imp}$ (see Eq.\ref{eq2}) as measured relatively to the reanalysis. The blue color means that the deep ensemble mean outperforms the free run and the red color means that it performs worse than the free run. The BC-RMSD's are calculated across the 2016-2020 period.}
\label{Fig.2}
\end{figure}

The comparison in Fig.\ref{Fig.1} is complemented by BC-RMSD$_{imp}$ metrics from Fig.\ref{Fig.2}, showing the \% improvement in BC-RMSD when comparing the BC-RMSD of the deep ensemble prediction using the reanalysis inputs with the BC-RMSD of the free run. The BC-RMSD is in both cases calculated separately for each spatial location and measured relative to the reanalysis data. Fig.\ref{Fig.2} clearly demonstrates that the deep ensemble prediction from reanalysis inputs substantially outperforms the free run in all surface carbon pools except for DIC (DOC could not be compared due to the lack of reanalysis output). This means for all the surface carbon pools except for DIC, it is preferable to use the deep ensemble prediction to running the free model simulation, and this is true not just for the time-averaged values (Fig.\ref{Fig.1}), but also for the time-series hidden behind the time-averages (Fig.\ref{Fig.2}). 
We did not identify any clear reason why the model fails to improve DIC relative to the free run, except that the biases between the free run and the reanalysis are substantially smaller for DIC than for the other variables (see Fig.\ref{Fig.0.5}-\ref{Fig.1}). In this context, it should be noted that the regions where the deep ensemble performs worse than the free run for DIC —the southern North Sea and coastal areas around the UK and the Irish Sea (Fig.~\ref{Fig.2}:D) — correspond to areas where the BC-RMSD of the free run relative to the reanalysis was found to be lowest, indicating its best performance (BC-RMSD $<$ 10 mmolC.m$^{-3}$; these results are not shown).
The comparatively small DIC biases in the free run can be attributed to several key drivers that are identical, or very similar, in both the free run and the reanalysis, including air–sea CO$_2$ fluxes, river discharge, and SST (for SST see Fig.\ref{Fig.0.5}). Consequently, given the relatively strong performance of the free run for DIC, the deep ensemble faces a substantially more stringent benchmark for further improvement.
Since, overall, the deep ensemble performs at a skill level comparable to that of the free run, the deep ensemble DIC product can be considered acceptable.

Curiously deep ensemble failed to improve the vertically averaged pools (not shown here). This can be understood based on SHapley Additive exPlanations (SHAP, \citet{lundberg2017unified}) analysis shown in Fig.A2-A3 of the Appendix. Using SHAPS it has been observed that for the vertically averaged variables the structural inputs (coordinates and bathymetry) are more important than they are for the surface variables (Fig.A2-A3 of the Appendix demonstrates this for heterotrophic bacteria carbon and for DOC). High importance of structural variables suggests that the deep ensemble mostly learned the free run climatology of the vertically averaged variables, showing too little flexibility when moving towards reanalysis. We have tried improving upon the structure of the NN models by including time-lagged features, to potentially represent longer time-scales associated with the vertically averaged variables, but there was no marked improvement to the results (not shown here). Finally, the SHAP analysis has shown consistently across all the carbon pools that the two most important groups of variables are oceanic inputs (SST, salinity, PFT chlorophyll) and structural variables, with atmospheric variables being less important and riverine discharge the least important.

\begin{figure}
\hspace{-1cm}
\noindent\includegraphics[width=14cm, height=11cm]{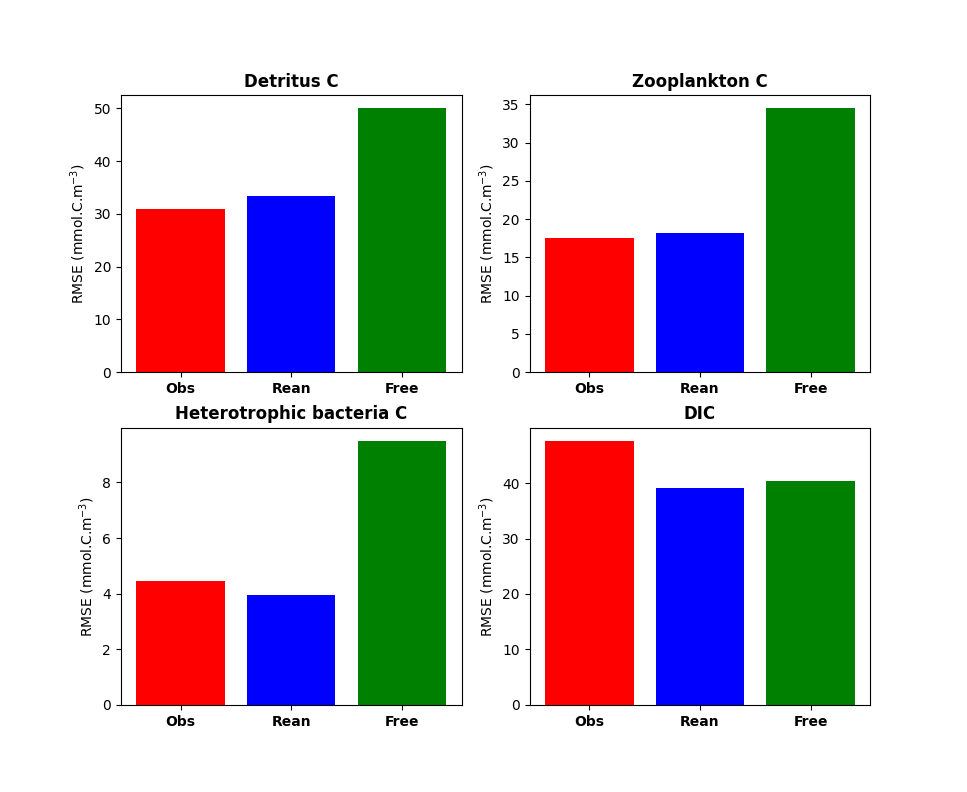}
\caption{RMSD skill (mmol.C/m$^{3}$) comparing performance of deep ensemble prediction of four surface carbon pools, using directly inputs from the assimilated observations (red), reanalysis (blue) and free run (green). The skill was evaluated against reanalysis carbon pools at the locations where satellite observations were available for the 2016-2020 period (amounting to over 50 000 data-points, see Sec.\ref{sec-exp}).}
\label{Fig.2.5}
\end{figure}

Although Fig.\ref{Fig.0.5}-\ref{Fig.1} show substantial differences between the reanalysis and the free run in both phytoplankton and the predicted carbon pools, one could argue that the reanalysis is not fully independent of the free run used to train the deep ensemble, since both are based on the same underlying physics–biogeochemistry model. To evaluate the skill of the deep ensemble directly against observations (being the ultimate proposed application of the deep ensemble), we therefore applied it to observational data. The validation, expressed in terms of RMSD (Fig.\ref{Fig.2.5}), indicates that the predictive skill for detritus, zooplankton, and heterotrophic bacteria carbon is comparable when using reanalysis and observational inputs, and substantially better than that of the free run. For DIC, the predictive skill obtained from observational inputs is slightly lower than for the other carbon pools; the underlying causes are likely multifactorial. Nevertheless, these results demonstrate that the deep ensemble is, for the most part, capable of predicting carbon pools from observational data with skill comparable to that achieved using reanalysis inputs (with the exception of DIC). This outcome is not entirely unexpected. Previous studies have shown that the biogeochemical analysis state in the Met Office NWES operational system exhibits close agreement with assimilated observations near the observation locations (e.g., \citet{skakala2018assimilation, skakala2019improved, skakala2020towards}). Although observational data are inherently noisier than reanalysis products, much of this noise is likely reduced through aggregation to the relatively coarse spatial and temporal resolution used for the NN inputs.

%- discuss that they can even explain much better the BC-RMSD

\begin{figure}
%\vspace{-2cm}
\hspace{1cm}
\vspace{0.5cm}
\noindent\includegraphics[width=10cm, height=15cm]{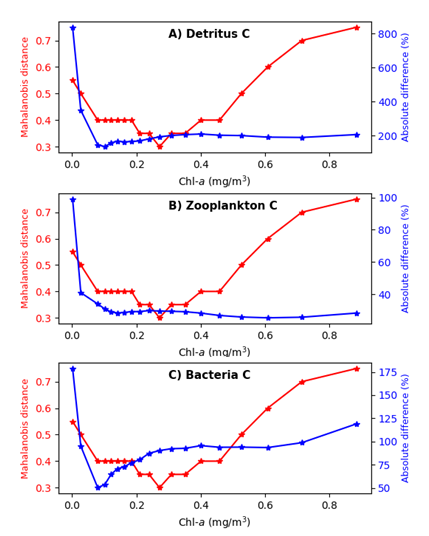}
%\vspace{-1cm}
\caption{The red line in each plot shows Mahalanobis distance (\citet{chandra1936generalised, ghorbani2019mahalanobis}) of the reanalysis inputs relative to the training and validation (training-validation) data calculated as follows: Chlorophyll-$a$ from the reanalysis (x-axis) was split into 20 quantiles and for each quantile we calculated median Mahalanobis distance between reanalysis and the training-validation data-set. Then we identified where (into which quantile) this median falls within the distribution of Mahalanobis distances within the training-validation data-set itself (calculated as distribution of Mahalanobis distances of the training-validation data from the training-validation data-set itself). The red y-axis then shows those quantiles on the 0-1 range of values, i.e. if the value is 0.5 it means that the median distance between reanalysis point and training-validation data-set was the same as the median distance among the training-validation data. If it is larger than 0.5, than the reanalysis points were generally further from the training-validation data-set than the median and vice versa. The blue line shows for the same quantiles the deep ensemble prediction skill measured against the reanalysis through mean absolute difference relative to the mean concentration (in \%).}
\label{Fig.3}
\end{figure}

There are two main reasons why the deep ensemble could have failed when predicting the carbon pools from the reanalysis (or observations): (i) the NN inputs provided by the reanalysis were too far from the data seen by the NN in the training process, (ii) the inherent relationship between the NN model inputs and the predicted NN outputs, learned by the NN, differs between the reanalysis and the model free run. 
Fig.~\ref{Fig.3} addresses point (i) by showing how Mahalanobis distances (\citet{chandra1936generalised, ghorbani2019mahalanobis}) between the reanalysis and the training-validation data vary across 20 reanalysis chlorophyll quantiles, and how they compare to Mahalanobis distances within the training–validation data set. The Figure also compares these distances with NN performance evaluated against the reanalysis.
%Fig.\ref{Fig.3} focuses on the point (i) by showing how Mahalanobis distances \citep{chandra1936generalised, ghorbani2019mahalanobis} between reanalysis and the training and validation (training-validation) data relate across 20 reanalysis chlorophyll quantiles to Mahalanobis distances within the training-validation data-set. The Figure further compares the Mahalanobis distance with the NN performance as measured against the reanalysis. 
Fig.\ref{Fig.3} shows that the reanalysis inputs were not far from what the deep ensemble has seen during training and validation, as their distance from the training and validation data-set was on average no significantly larger than the typical distance between the data-points within the training and validation data-set. 

\begin{figure}
\hspace{-1.7cm}
\noindent\includegraphics[width=16cm, height=11cm]{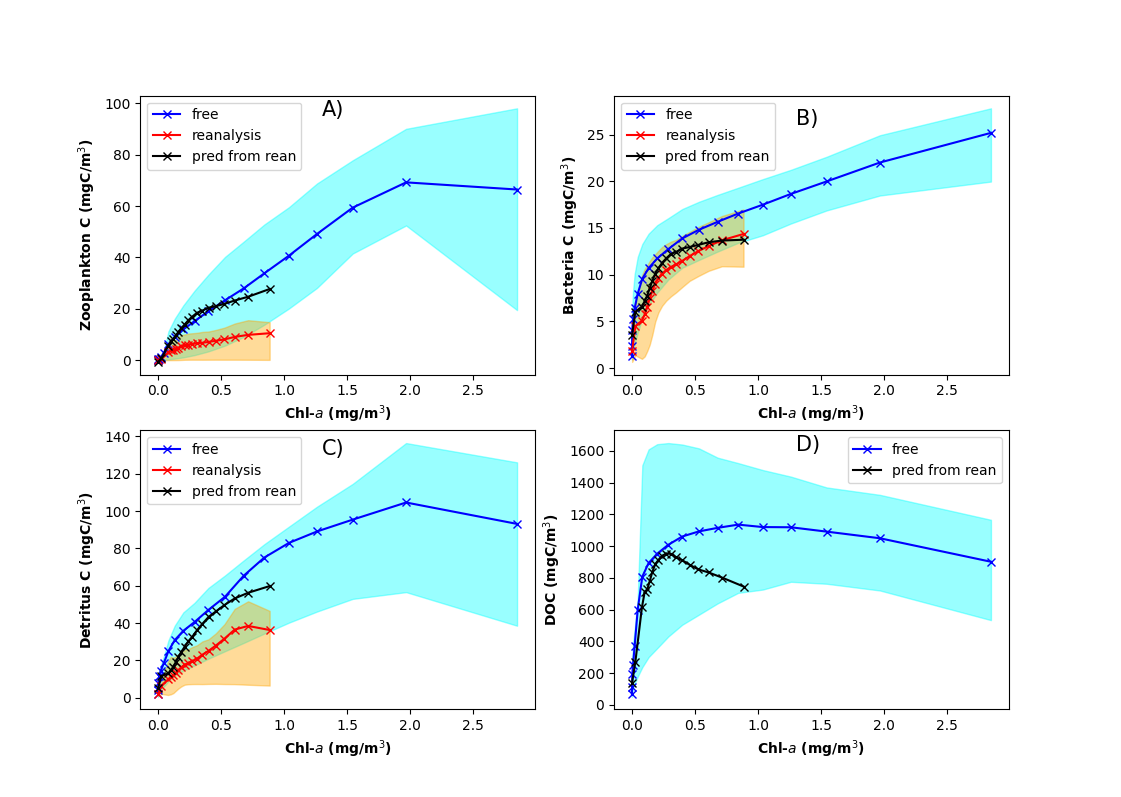}
\caption{On the x-axis we show phytoplankton surface chlorophyll-$a$ concentrations (in mg/m$^{3}$). The markers on the lines show on the x-axis 20 quantiles into which the total chlorophyll-$a$ distribution (aggregated across spatial domain and 2016-2020 period) was split. On the y-axis the markers show for each chlorophyll-$a$ quantile the median values of the corresponding carbon pools: zooplankton (A), heterotrophic bacteria (B), detritus (C) and DOC (D). The shaded colors show the spread of the values for each carbon pool across each chlorophyll-$a$ quantile, i.e. the spread is defined for each carbon pool as two quartiles around its median value for each chlorophyll-$a$ quantile. The differently colored lines and shaded areas account for the model free run (blue), reanalysis (red) and the deep ensemble prediction using reanalysis as inputs (black). For the deep ensemble prediction only the median carbon pools are shown (no shaded areas).}
\label{Fig.4}
\end{figure}

Addressing the point (ii) is a little bit more tricky: although the relationship between NN model inputs and the predicted outputs in the reanalysis is provided via the dynamical adjustment of the same biogeochemistry model, whose free run was used to train the NN, there is no guarantee that the map from the NN inputs to the outputs within the reanalysis will not get distorted by the impact of assimilation. An interesting insight is provided by Fig.\ref{Fig.4} showing the dependence of surface carbon pools on chlorophyll-$a$. Although the dependency plotted in Fig.\ref{Fig.4} is a major simplification of the true function learned by the deep ensemble, it indicates that the functions between NN inputs and detritus and zooplankton carbon pools might significantly differ between the model free run and the reanalysis. It is also notable that for zooplankton and detritus the deep ensemble predicts similar functional dependence as the free run. This suggests that the deviation between the deep ensemble prediction and the reanalysis in Fig.\ref{Fig.4}:A,C is due to differences in those functional dependencies between the free run and the reanalysis, i.e. deep ensemble using the function it learned in the free run to create predictions on the space of reduced chlorophyll-$a$ concentrations from the reanalysis. Interestingly the same is less true for the heterotrophic bacteria carbon and the DOC (Fig.\ref{Fig.4}:B,D).

%This interpretation is supported by the fact that the dependence of the surface carbon pools on chlorophyll-$a$ predicted by the NN model is at least for zooplankton and detritus very close to the dependence found in the training and validation data.

%- SHAP plot and explainability discussion

%- Explain in terms of phyto vs pools C dependence - two sources that can limit the ML performance is distance from the training data and difference in the statistical relationship due to spatio-temporal dependence, difference in model etc. The latter is not expected for reanalysis, as both cross-covariances and dynamical adjustment depend on the same model that is used to train the ML. It is however certainly relevant for climate what-if scenarios. Fig.5-6 show that the reanalysis data are in reasonable neighborhood of training data. 

\begin{figure}
\vspace{-4cm}
\hspace{-2.7cm}
\noindent\includegraphics[width=18cm, height=20cm]{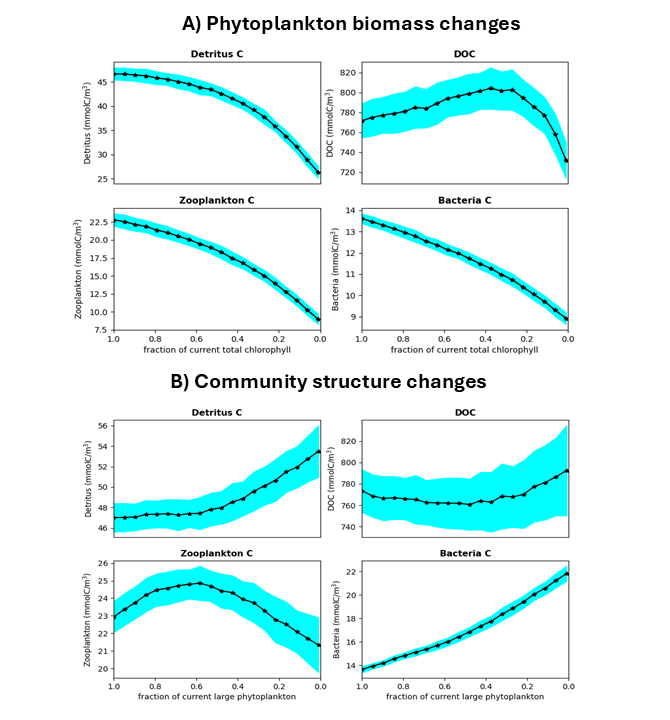}
\caption{A) Deep ensemble predicted what-if scenario for how decline in phytoplankton biomass measured as a fraction of its present 2016-2020 reanalysis value (on [0-1] scale) maps into the estimated carbon pools. B) Similar what-if scenario for changes in the community structure measured by the ratio in chlorophyll-$a$ biomass between macrophytoplankton (diatoms and dinoflaggelates) and total phytoplankton, relative to the average reanalysis 2016-2020 ratio (again on a [0-1] scale). The shaded areas show in both cases the uncertainty derived from the deep ensemble standard deviation.}
\label{Fig.5}
\end{figure}

It is possible to use the lightweight NN models to simulate a wide range of what-if scenarios, something hard to do with computationally expensive process-based models. The future climate projections for the NWES indicate important changes to a range of variables included as inputs in the NN (\citet{wakelin2015modelling}). This includes decline in near-surface primary production and therefore phytoplankton biomass, and changes to phytoplankton community structure, with increased proportion of smaller size-classes in the phytoplankton community. Both these changes are thought to be primarily driven through the increased thermal stratification of the ocean, cutting off the nutrients from the ocean surface (\citet{wakelin2015modelling}). In Fig.\ref{Fig.5} we plot the response of ocean surface carbon pools as predicted by the deep ensemble to such changes in the surface phytoplankton (i.e. ``what-if'' means here ``what happens with ocean carbon if such changes occur''). The functional relationships from Fig.\ref{Fig.5}:A can be mostly easily understood (except for DOC), as they just indicate an overall decline of organic carbon pools with the decline of phytoplankton biomass. However, the dependence of the same ocean surface carbon pools on the ocean surface phytoplankton community structure (Fig.\ref{Fig.5}:B) is in some cases more complex. Whilst the surface detrital carbon monotonically increases with reduced large-size phytoplankton presence likely due to decrease in sinking rates and carbon export, the non-monotonous changes in DOC and zooplankton carbon pools could be influenced by the changes in their own functional type compositions as a function of the modified phytoplankton community structure. It should be noted, however, that the limitations of the NN-based approach are immediately apparent: a decline of phytoplankton biomass to zero (Fig.~\ref{Fig.5}:A) does not lead to a corresponding reduction of the carbon pools to zero. This indicates that the NN is operating far outside the range over which it was trained and, at some point, begins to underestimate ecosystem changes associated with declining phytoplankton biomass.

\begin{figure}
\hspace{-1.5cm}
\noindent\includegraphics[width=16cm, height=10cm]{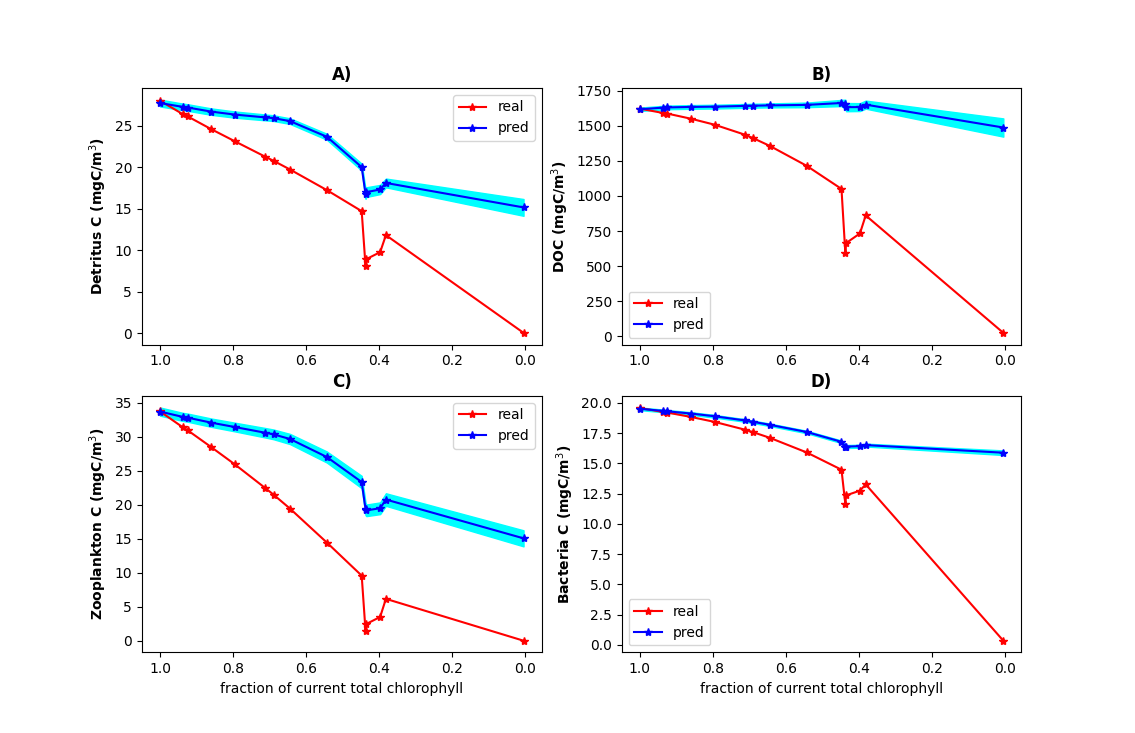}
%\makeatletter 
%\renewcommand{\thefigure}{A4}%\@arabic\c@figure}
%\makeatother
\caption{1D experiments forced by nutrient relaxation to a state with reduced surface phytoplankton concentrations representing changed climate. In red is the relationship between time-averaged total chlorophyll-a (x-axis) and time-averaged carbon pools (y-axis) from the 1D model runs. Each 1D simulation was run for 20 years with 5 years used as spin-up and 15 years used to calculate the averages for each "climate" state. The simulations were run at L4 location in the western English Channel with atmospheric forcing taken from real 2001-2020 data (hence the only forcing changing the climate state was the nutrient relaxation). In blue is deep ensemble prediction of the surface carbon pools, using the inputs from the 1D runs. The deep ensemble was trained on the simulation with the present state nutrients. The Figure indicates in the simplified 1D setting how much is the deep ensemble capable to generalize from the current climate to different climate states.}
\label{Fig.6}
\end{figure}

There are many reasons to be cautious about the results presented in Fig.\ref{Fig.5}. Even though the what-if scenario input features did not have anomalously large Mahalanobis distance from the training data (not shown here), one can raise serious doubts about how the statistical relationship learned by the deep ensemble in the model hindcast translates into a hypothetical future climate. We have tested this by a series of 1D simulations using Generalised Ocean Turbulence Model (GOTM, \citet{burchard1999gotm}) coupled to ERSEM through FABM. In those simulations we reduced nutrients by relaxing them towards lowered climatology values, triggering decrease in primary production and phytoplankton in the 1D model. We have trained a deep ensemble based on the simulation data where nutrients were not reduced and we analysed how well the deep ensemble predicts the carbon pools from the 1D simulations with reduced nutrients. These experiments (see Fig.\ref{Fig.6}) demonstrate that the deep ensemble systematically underestimates the effects of reduced phytoplankton on carbon pools, with the magnitude of this bias increasing as phytoplankton biomass declines. The 1D experiments would then nicely explain why the carbon pools in Fig.\ref{Fig.5} do not approach zero as the phytoplankton concentration vanishes. However, the deep ensemble was capable to do a decent job in estimating bacteria carbon changes until the phytoplankton biomass was reduced by more than 30\% (Fig.\ref{Fig.6}:D). This may be explained by the comparatively weaker response of bacterial concentrations to reductions in phytoplankton biomass, relative to the responses of the other carbon pools (at least until phytoplankton declines to approximately half of its baseline concentration; see Fig.~\ref{Fig.6}:D). The relative stability of bacterial carbon under changing phytoplankton climatology would therefore make it more predictable by an ML model trained on the present-day climate simulation. An interesting feature is the sudden dip in carbon pools relative to phytoplankton biomass reduction around 40\% of its present value (Fig.\ref{Fig.6}). We have observed (not shown) that this is due to a sudden change in phytoplankton composition (increase in large species) in such a climate, and interestingly consequences of this change in phytoplankton composition for the carbon pools are broadly correctly predicted by the NN (similar dip is predicted by the NN). Overall, this simple exercise shows that although similar what-if scenarios using deep ensembles should be taken with healthy suspicion, there might be specific cases where the prediction works well within a relatively wide range of scenarios.

Finally, we took initial steps towards estimating the uncertainty in the marine physics–\-biogeochemistry model–guided, ML-learned function that maps observable variables to carbon pools. The uncertainty in this function is expected to arise mainly from two sources: (i) model structural deficiencies (e.g., errors in the structural form of equations or unresolved variability) and (ii) highly uncertain model parameter values (for discussion of this in the context of ERSEM, see \citet{skakala2024how}). Addressing structural deficiencies was beyond the scope of this study; however, we performed initial analyses to estimate the impact of parameter uncertainty. To this end, we used the same 1D model configuration at the L4 location as in the climate scenario experiments, but ran a 100-member ensemble in which five ERSEM model parameters (such as diatom maximum productivity at reference temperature, or the diatom maximum nitrogen-to-carbon ratio) were perturbed. These parameters were selected based on the sensitivity analysis of \cite{ciavatta2025control}. The ensemble simulations were conducted using the Ensemble and Assimilation Tool described by \citet{bruggeman2023eat}. The five parameters were sampled from uniform distributions within $\pm$30\% of their reference values. One hundred different ML models were then trained using outputs from the 100 ensemble members, which differed only in their parameter values. The 100 ML models were subsequently applied to the same input data, and we evaluated the spread in their predictions of carbon pools. In this simplified setting, prior uncertainty in the model parameters resulted in a typical standard deviation of surface organic carbon pools of approximately 30–40\% of their mean value (not shown). We recommend extending this highly simplified 1D analysis in future work to obtain a more realistic assessment of uncertainty in the ML models.

%- speculative what-if scenarios, curiously the Mahalanbis distance from training data is not too bad, but clearly the statistics will be dependent on changing climate. How much is shown by the 1D experiment

%\begin{figure}
%\hspace{-2cm}
%\noindent\includegraphics[width=16cm, height=10cm]{what_if_comms_with_uncertainty.png}
%\caption{}
%\label{Fig.8}
%\end{figure}

\section{Conclusions and the next steps}

In this work, we demonstrate that a variety of surface carbon pools (detrital, dissolved organic carbon, zooplankton, heterotrophic bacteria carbon and dissolved inorganic carbon - DIC) in the Northwest European Shelf (NWES) can be estimated (on 35 km and 10-day scale) using model-informed machine learning (ML). This approach has the potential to serve as a computationally highly efficient substitute for reanalysis systems by predicting the carbon pools directly from atmospheric, riverine data and satellite observations. The ML model, based on a deep ensemble of relatively simple and easy-to-train perceptron neural networks, can generate predictions spanning multiple years of data within seconds on a desktop computer. In contrast, reanalysis requires approximately one hour of wall-clock time on a supercomputer using O(100) cores to simulate a single day. However, reanalysis retains a key advantage in its ability to provide more detailed outputs in terms of both variables, spatial and temporal coverage and resolution. The extent to which this level of detail can be reproduced by ML-based approaches remains an open question.

Although the deep ensemble enables rapid predictions, it may require substantial resources for generating training data. In our case, these data are readily available from existing, freely accessible coupled marine physics–biogeochemistry model simulations. The trained deep ensemble was then applied to estimate surface carbon pools (along with their associated uncertainties) directly from observations of physical variables and ocean colour–derived phytoplankton functional type (PFT) chlorophyll, as well as atmospheric, riverine, and structural data.
%In this work we have demonstrated that we can estimate variety of surface carbon pools (detrital, DOC, zooplankton and heterotrophic bacteria carbon) in the NWES using model-informed machine learning (ML), which could effectively replace reanalysis by a highly cost-effective emulator. The ML model consisting of deep ensemble of relatively simple and easy-to-train perceptron neural networks provides predictions for multiple years of data on the order of seconds on a desktop computer, whereas the reanalysis requires for a single simulation day about one hour of wall-clock time on a supercomputer using O(100) cores. The advantage of reanalysis however remains in providing much more detailed outputs in terms of both variables and resolution, and it remains only to be seen how much of this capability can be reproduced with ML. 
%Even though the deep ensemble produces fast predictions, it might need substantial resources in terms of generating training data. In our application these are however readily available in the existing free coupled marine physics-biogeochemistry model runs. The deep ensemble trained on the free run data was then applied (on a 35km spatial and 10-daily temporal scales) to estimate the surface carbon pools (and their uncertainties) directly from the observations of physics variables and ocean color-derived PFT chlorophyll, as well as atmospheric, riverine and structural data. 
We have shown that the deep ensemble agrees much better with the reanalysis than the free run for all predicted surface pools where the free run exhibits large biases relative to the reanalysis (i.e., all except DIC). 

Such model-informed ML could complement the existing satellite algorithms for some of the carbon pools in regions where they perform poorly, or provide carbon pools where such satellite algorithms are entirely absent. For example, in the case of NWES there are no satellite-derived products for detrital, zooplankton and heterotrophic bacteria carbon (and very little in situ data as well), so the reanalysis assimilating physics data and ocean color PFT chlorophyll-$a$ is taken as the best available estimate of those pools. Wherever the existing spatial and temporal coverage by the satellite data is considered sufficient, we propose to use the ML model developed in this study as an alternative to the relatively computationally costly NWES reanalysis, or a real-time monitoring system. Furthermore, we recommend considering similar tools also in different regions. The degree to which the ML model developed here for the NWES is transferrable to those different regions remains to be determined in the future work. If transferability is an issue, the development of NN models will depend on the availability of model simulations for those new regions, or it will incur additional costs to construct a new training dataset.

As the surface carbon pools carry only limited information about carbon cycle, estimating vertical profiles of those pools is quite essential, e.g. to get answers to questions such as how much carbon is exported into the deep ocean. The approach adopted in this work was able to only learn the seasonal climatology of the vertically averaged carbon pools. Although there are likely hard limits on how much one can dynamically predict vertically averaged pools from the surface inputs, the relative simplicity of the ML architecture adopted here might have been a contributing factor to ML capturing only the seasonal climatology of the vertically averaged pools. Even though implementing time-lagged features did not seem to help, there is a chance that improvement can be made with more sophisticated models and longer data-sets. As part of increasing model complexity, one could also consider replacing the structural ML inputs with new flow-dependent features, e.g. capturing the underlying hydrodynamics. Such new features would however likely require using observations from profiles, significantly constraining the spatial domain where the ML model can be applied, or a 3D physical model run, making the applicability of the ML model significantly more demanding in terms of the required inputs into the model (relying on 3D hydrodynamic model runs).

Finally, other applications of the existing ML surface carbon prediction can be imagined, these include implementing assimilation of the ML-derived surface carbon alongside the currently assimilated variables, similarly to what was done with nitrate in \citet{banerjee2025combining}. If this was done ``online'' with the ML prediction cycled with the DA (see the discussion in \citet{banerjee2025combining}) we could improve the speed with which the model adjusts itself to assimilation of standardly provided satellite data (such as OC chlorophyll-$a$). We suggest to explore these routes in the future.

\vspace{1cm}

%We believe that the model-informed ML can prove to be beneficial also for other purposes, such as DA balancing...

%{\bf Acknowledgments:} This work was funded by the Horizon Europe project The New Copernicus Capability
%for Tropic Ocean Networks (NECCTON, grant agreement no.101081273). We also acknowledge support from the UK Natural Environment Research Council, including the single centre national capability programme – Climate Linked Atlantic Sector Science (CLASS, 379NE/R015953/1). I would like to thank David Ford and Richard Renshaw for providing me with access to the Copernicus reanalysis data on the Met Office HPC MonSOON2 storage system MASS.

{\bf Acknowledgments:} This work was funded by the Horizon Europe project The New Copernicus Capability
for Tropic Ocean Networks (NECCTON, grant agreement no.101081273).  I also acknowledge financial support from the UK Natural Environment Research Council, including the single centre national capability programme—Climate Linked Atlantic Sector Science (Climate Linked Atlantic Sector Science, 379NE/R015953/1), Atlantic Climate and Environment Strategic Science (Atlantis), and National Centre for Earth Observation (NCEO). I would like to thank David Ford and Richard Renshaw for providing me with access to the Copernicus reanalysis data on the Met Office HPC MonSOON2 storage system MASS. I would also like to thank Helen Powley and David Moffat for discussions and Deep S. Banerjee for providing me with processed river data inputs. 

{\bf Conflict of interest:} The author declares no conflict of interest relevant to this study.

{\bf Open Research:} The simulation and reanalysis data are all stored and available through the Met Office MASS system. The full reanalysis can be downloaded under ID: rosie\_\-u-bn555\_\-jan98\_\-pft3.1\_\-bias\_\-r133976\_\-frdf and partly can be obtained from https://\-data.\-marine.\-copernicus.eu/\-product/\-NWSHELF\_\-MULTIYEAR\_\-BGC\_\-004\_\-011/description (\citet{kay2016north-data}). The processed version of the 2016-2020 reanalysis can be downloaded from \citet{skakala2025ecosystem-data2}. The processed version of the free run can be downloaded from \citet{skakala2025ecosystem-data}, or obtained in full under ID: u-ci193 (for instructions please see https://\-help.\-jasmin.\-ac.uk/\-docs/\-mass/). The NN model and the related code (\citet{skakala2025ecosystem-software}) used in this study can be obtained through Zenodo on https://doi.org/10.5281/\-zenodo.20276280.

\bibliography{Carbon_Skakala_ArXiv}

@article{artioli2012carbonate,
  title={The carbonate system in the North Sea: Sensitivity and model validation},
  author={Artioli, Yuri and Blackford, Jeremy C and Butensch{\"o}n, Momme and Holt, Jason T and Wakelin, Sarah L and Thomas, Helmuth and Borges, Alberto and Allen, J Icarus},
  journal={Journal of Marine Systems},
  volume={102},
  pages={1--13},
  year={2012},
  publisher={Elsevier}
}

@article{stramski2008relationships,
  title={Relationships between the surface concentration of particulate organic carbon and optical properties in the eastern South Pacific and eastern Atlantic Oceans},
  author={Stramski, Dariusz and Reynolds, Rick A and Babin, Marcel and Kaczmarek, S and Lewis, Marlon R and R{\"o}ttgers, R and Sciandra, Antoine and Stramska, M and Twardowski, MS and Franz, BA and Claustre, H.},
  journal={Biogeosciences},
  volume={5},
  number={1},
  pages={171--201},
  year={2008},
  publisher={Copernicus GmbH}
}

@article{neccton2025,
  title={NECCTON: Technical specification of the pelagic lower trophic level products},
  author={Clark, James and Kay, Susan and Samuelsen, Annette and Conchon, Anna and Albernhe, Sarah and Butenschon, Momme and
Powley, Helen and Cossarini, Gianpiero and Lazzari, Paolo OGS and Hahn, Josephine and Braga, Frederica and Perruche, Corralie and Gregoire, Marilaure},
  journal={Zenodo},
  year={2025},
  doi={https://doi.org/10.5281/zenodo.10057818}
}

@article{wakelin2015modelling,
  title={Modelling the combined impacts of climate change and direct anthropogenic drivers on the ecosystem of the northwest European continental shelf},
  author={Wakelin, Sarah and Artioli, Yuri and Butensch{\"o}n, Momme and Allen, J Icarus and Holt, Jason T},
  journal={Journal of Marine Systems},
  volume={152},
  pages={51--63},
  year={2015},
  publisher={Elsevier}
}

@article{banerjee2025combining,
  title={Combining machine learning with data assimilation to improve the quality of phytoplankton forecasting in a shelf sea environment},
  author={Banerjee, Deep and Skakala, Jozef and Ford, David},
  journal={submitted to QJRMS, arXiv:2508.02400, https://arxiv.org/abs/2508.02400},
  year={2025}
}

@article{mannino2008algorithm,
  title={Algorithm development and validation for satellite-derived distributions of DOC and CDOM in the US Middle Atlantic Bight},
  author={Mannino, Antonio and Russ, Mary E and Hooker, Stanford B},
  journal={Journal of Geophysical Research: Oceans},
  volume={113},
  number={C7},
  year={2008},
  publisher={Wiley Online Library}
}

@article{brewin2017uncertainty,
title={Uncertainty in ocean-color estimates of chlorophyll for phytoplankton groups},
author={Brewin, Robert and Ciavatta, Stefano and Sathyendranath, Shubha and Jackson, Tom and Tilstone, Gavin and Curran, Kieran and Airs, Ruth and Cummings, Denise and Brotas, Vanda and Organelli, Emmanuele and Dall'Olmo, Giorgio},
journal={Frontiers in Marine Science},
pages={104},
year={2017}
}

@article{fablet2021joint,
title={Joint interpolation and representation learning for irregularly sampled satellite-derived geophysical fields},
author={Fablet, R. and Beauchamp, M. and Drumetz, L. and Rousseau, F.},
journal={Frontiers in Applied Mathematics and Statistics},
volume={7},
pages={655224},
year={2021}
}

@article{wakelin2012modeling,
  title={Modeling the carbon fluxes of the northwest European continental shelf: Validation and budgets},
  author={Wakelin, Sarah and Holt, Jason and Blackford, Jerry and Allen, Icarus and Butensch{\"o}n, Momme and Artioli, Yuri},
  journal={Journal of Geophysical Research: Oceans},
  volume={117},
  number={C5},
  year={2012},
  publisher={Wiley Online Library}
}

@article{higgs2023ecosystem,
  title={Investigating ecosystem connections in the shelf sea environment using complex networks},
  author={Higgs, Ieuan and Sk{\'a}kala, Jozef and Bannister, Ross and Carrassi, Alberto and Ciavatta, Stefano},
  journal={Biogeosciences},
  volume={21},
  pages={731--746},
  year={2024},
  publisher={Copernicus Publications G{\"o}ttingen, Germany}
}

@dataset{skakala2025ecosystem-data,
  author    = {Sk{\'a}kala, Jozef},
  title     = {NEMO-FABM-ERSEM free run 2016-2020 simulations},
  year      = {2025},
  publisher = {[Dataset], GitHub},
  url       = {https://github.com/JOZSKA/ML_for_carbon_pools}
}

@dataset{skakala2025ecosystem-data2,
  author    = {Sk{\'a}kala, Jozef},
  title     = {NEMO-FABM-ERSEM reanalysis 2016-2020 simulations},
  year      = {2025},
  publisher = {[Dataset], GitHub},
  url       = {https://github.com/JOZSKA/ML_for_carbon_pools}
}

@software{skakala2025ecosystem-software,
  author    = {Sk{\'a}kala, Jozef},
  title     = {NN code and weights},
  year      = {2025},
  publisher = {[Software], GitHub},
  url       = {https://github.com/JOZSKA/ML_for_carbon_pools}
}

@article{friedlingstein2024global,
  title={Global carbon budget 2024},
  author={Friedlingstein, Pierre and O'sullivan, Michael and Jones, Matthew W and Andrew, Robbie M and Hauck, Judith and Landsch{\"u}tzer, Peter and Le Qu{\'e}r{\'e}, Corinne and Li, Hongmei and Luijkx, Ingrid T and Olsen, Are and Zeng, Jiye},
  journal={Earth System Science Data Discussions},
  volume={2024},
  pages={1--133},
  year={2024},
  publisher={G{\"o}ttingen, Germany}
}

@article{borges2006carbon,
  title={Carbon dioxide in European coastal waters},
  author={Borges, AV and Schiettecatte, L-S and Abril, Gwena{\"e}l and Delille, Bruno and Gazeau, Fr{\'e}d{\'e}ric},
  journal={Estuarine, coastal and shelf science},
  volume={70},
  number={3},
  pages={375--387},
  year={2006},
  publisher={Elsevier}
}

@incollection{jahnke2010global,
  title={Global synthesis1},
  author={Jahnke, Richard A},
  booktitle={Carbon and nutrient fluxes in continental margins: A global synthesis},
  pages={597--615},
  year={2010},
  publisher={Springer}
}

@book{emerson2008chemical,
  title={Chemical oceanography and the marine carbon cycle},
  author={Emerson, Steven and Hedges, John},
  year={2008},
  publisher={Cambridge University Press}
}

@article{volk1985ocean,
  title={Ocean carbon pumps: Analysis of relative strengths and efficiencies in ocean-driven atmospheric CO2 changes},
  author={Volk, Tyler and Hoffert, Martin I},
  journal={The carbon cycle and atmospheric CO2: Natural variations Archean to present},
  volume={32},
  pages={99--110},
  year={1985},
  publisher={Wiley Online Library}
}

@article{groom2019satellite,
  title={Satellite ocean colour: current status and future perspective},
  author={Groom, Steve and Sathyendranath, Shubha and Ban, Yai and Bernard, Stewart and Brewin, Robert and Brotas, Vanda and Brockmann, Carsten and Chauhan, Prakash and Choi, Jong-kuk and Chuprin, Andrei and Wang, Menghua},
  journal={Frontiers in Marine Science},
  volume={6},
  pages={485},
  year={2019},
  publisher={Frontiers Media SA}
}

@article{brewin2021sensing,
  title={Sensing the ocean biological carbon pump from space: A review of capabilities, concepts, research gaps and future developments},
  author={Brewin, Robert and Sathyendranath, Shubha and Platt, Trevor and Bouman, Heather and Ciavatta, Stefano and Dall'Olmo, Giorgio and Dingle, James and Groom, Steve and J{\"o}nsson, Bror and Kostadinov, Tihomir S and Walker, Peter},
  journal={Earth-Science Reviews},
  volume={217},
  pages={103604},
  year={2021},
  publisher={Elsevier}
}

@article{evers2017validation,
  title={Validation and intercomparison of ocean color algorithms for estimating particulate organic carbon in the oceans},
  author={Evers-King, Hayley and Martinez-Vicente, Victor and Brewin, Robert JW and Dall'Olmo, Giorgio and Hickman, Anna E and Jackson, Thomas and Kostadinov, Tihomir S and Krasemann, Hajo and Loisel, Hubert and R{\"o}ttgers, R{\"u}diger and Sathyendranath, Shubha},
  journal={Frontiers in Marine Science},
  volume={4},
  pages={251},
  year={2017},
  publisher={Frontiers Media SA}
}

@article{le2018color,
  title={A color-index-based empirical algorithm for determining particulate organic carbon concentration in the ocean from satellite observations},
  author={Le, Chengfeng and Zhou, Xueying and Hu, Chuanmin and Lee, Zhongping and Li, Lin and Stramski, Dariusz},
  journal={Journal of Geophysical Research: Oceans},
  volume={123},
  number={10},
  pages={7407--7419},
  year={2018},
  publisher={Wiley Online Library}
}

@article{roy2017size,
  title={Size-partitioned phytoplankton carbon and carbon-to-chlorophyll ratio from ocean colour by an absorption-based bio-optical algorithm},
  author={Roy, Shovonlal and Sathyendranath, Shubha and Platt, Trevor},
  journal={Remote Sensing of Environment},
  volume={194},
  pages={177--189},
  year={2017},
  publisher={Elsevier}
}

@article{sathyendranath2020reconciling,
  title={Reconciling models of primary production and photoacclimation},
  author={Sathyendranath, Shubha and Platt, Trevor and Kova{\v{c}}, {\v{Z}}arko and Dingle, James and Jackson, Thomas and Brewin, Robert JW and Franks, Peter and Mara{\~n}{\'o}n, Emilio and Kulk, Gemma and Bouman, Heather A},
  journal={Applied Optics},
  volume={59},
  number={10},
  pages={C100--C114},
  year={2020},
  publisher={OSA}
}

@article{lundberg2017unified,
  title={A unified approach to interpreting model predictions},
  author={Lundberg, Scott M and Lee, Su-In},
  journal={Advances in neural information processing systems},
  volume={30},
  year={2017}
}

@article{hochreiter1997long,
  title={Long short-term memory},
  author={Hochreiter, Sepp and Schmidhuber, J{\"u}rgen},
  journal={Neural computation},
  volume={9},
  number={8},
  pages={1735--1780},
  year={1997},
  publisher={MIT press}
}

@inproceedings{chandra1936generalised,
  title={On the generalised distance in statistics},
  author={Chandra, Mahalanobis Prasanta},
  booktitle={Proceedings of the National Institute of Sciences of India},
  volume={2},
  number={1},
  pages={49--55},
  year={1936}
}

@article{ghorbani2019mahalanobis,
  title={Mahalanobis distance and its application for detecting multivariate outliers},
  author={Ghorbani, Hamid},
  journal={Facta Universitatis, Series: Mathematics and Informatics},
  pages={583--595},
  year={2019}
}

@article{yu2019review,
  title={A review of recurrent neural networks: LSTM cells and network architectures},
  author={Yu, Yong and Si, Xiaosheng and Hu, Changhua and Zhang, Jianxun},
  journal={Neural computation},
  volume={31},
  number={7},
  pages={1235--1270},
  publisher={MIT Press One Rogers Street, Cambridge, MA 02142-1209, USA journals-info~…},
  year={2019}
}

@article{li2021survey,
  title={A survey of convolutional neural networks: analysis, applications, and prospects},
  author={Li, Zewen and Liu, Fan and Yang, Wenjie and Peng, Shouheng and Zhou, Jun},
  journal={IEEE transactions on neural networks and learning systems},
  volume={33},
  number={12},
  pages={6999--7019},
  year={2021},
  publisher={IEEE}
}

@article{matsuoka2017pan,
  title={Pan-Arctic optical characteristics of colored dissolved organic matter: Tracing dissolved organic carbon in changing Arctic waters using satellite ocean color data},
  author={Matsuoka, Atsushi and Boss, Emmanuel and Babin, Marcel and Karp-Boss, Lee and Hafez, Mark and Chekalyuk, Alex and Proctor, Christopher W and Werdell, P Jeremy and Bricaud, Annick},
  journal={Remote sensing of Environment},
  volume={200},
  pages={89--101},
  year={2017},
  publisher={Elsevier}
}

@article{laine2024machine,
  title={A machine learning model-based satellite data record of dissolved organic carbon concentration in surface waters of the global open ocean},
  author={Laine, Marko and Kulk, Gemma and J{\"o}nsson, Bror F and Sathyendranath, Shubha},
  journal={Frontiers in Marine Science},
  volume={11},
  pages={1305050},
  year={2024},
  publisher={Frontiers Media SA}
}

@article{maclachlan2024global,
title={Global seasonal forecast system version 5 (GloSea5): A high‐resolution seasonal forecast system},
author={MacLachlan, C. and Arribas, A. and Peterson, K.A. and Maidens, A. and Fereday, D. and Scaife, A.A. and Gordon, M. and Vellinga, M. and Williams, A. and Comer, R.E. and Camp, J.}, 
journal={Quarterly Journal of the Royal Meteorological Society},
volume={141},
pages={1072-1084},
year={2015}
}

@article{scarselli2008the,
title={The graph neural network model},
author={Scarselli, F. and Gori, M. and Tsoi, A.C. and Hagenbuchner, M. and Monfardini, G.},
journal={IEEE transactions on neural networks},
volume={20},
pages={61-80},
year={2008}
}

@article{zhou2020graph,
title={Graph neural networks: A review of methods and applications},
author={Zhou, J. and Cui, G. and Hu, S. and Zhang, Z. and Yang, C. and Liu, Z. and Wang, L. and Li, C. and Sun, M.},
journal={AI Open},
volume={1},
pages={57-81},
year={2020}
}

@article{li2024advanced, 
title={Advanced Machine Learning Models for Estimating the Distribution of Sea-Surface Particulate Organic Carbon (POC) Concentrations Using Satellite Remote Sensing Data: The Mediterranean as an Example},
author={Li, C. and Wu, H. and Yang, C. and Cui, L. and Ma, Z. and Wang, L.},
journal={Sensors},
volume={24},
pages={5669},
year={2024}
}

@article{sauzede2020estimation, 
title={Estimation of oceanic particulate organic carbon with machine learning},
author={Sauzède, R. and Johnson, J. and Claustre, H. and Camps-Valls, G. and Ruescas, A.},
journal={ISPRS Annals of Photogrammetry},
volume={2},
pages={949-956},
year={2020}
}

@article{zhang2025review,
title={A review of machine learning applications in ocean color remote sensing},
author={Zhang, Z. and Chen, P. and Zhang, S. and Huang, H. and Pan, Y. and Pan, D.},
journal={Remote Sensing},
volume={17},
pages={1776},
year={2025}
}

@article{ciavatta2025control,
title={Control of simulated ocean ecosystem indicators by biogeochemical observations},
author={Ciavatta, S. and Lazzari, P. and Alvarez, E. and Bertino, L. and Bolding, K. and Bruggeman, J. and Capet, A. and Cossarini, G. and Daryabor, F. and Nerger, L. and Popov, M., and Skakala, J. and Spada, S. and Teruzzi, A. and Wakamatsu, T. and Yumruktepe, V.Ç. and Brasseur, P.},
journal={Progress in Oceanography},
volume={231},
pages={103384},
year={2025}
}

@article{raissi2019a, 
title={Physics-informed neural networks: A deep learning framework for solving forward and inverse problems involving nonlinear partial differential equations},
author={Raissi, M. and Perdikaris, P. and Karniadakis, G.E.}, 
journal={Journal of Computational physics},
volume={378},
pages={686-707},
year={2019}
}

@article{bruggeman2023eat, 
title={Eat v0. 9.6: a 1d testbed for physical-biogeochemical data assimilation in natural waters},
author={Bruggeman, J. and Bolding, K. and Nerger, L. and Teruzzi, A. and Spada, S. and Skákala, J. and Ciavatta, S.},
journal={Geoscientific Model Development},
pages={5619–5639},
volume={17},
year={2024}
}

@article{zemskova2022deep, 
title={A deep-learning estimate of the decadal trends in the Southern Ocean carbon storage},
author={Zemskova, V.E. and He, T.L. and Wan, Z. and Grisouard, N.},
journal={Nature Communications},
volume={13},
pages={4056},
year={2022}
}

@article{stromberg2009estimation,
  title={Estimation of global zooplankton biomass from satellite ocean colour},
  author={Str{\"o}mberg, KH Patrik and Smyth, Timothy J and Allen, J Icarus and Pitois, Sophie and O'Brien, Todd D},
  journal={Journal of Marine Systems},
  volume={78},
  number={1},
  pages={18--27},
  year={2009},
  publisher={Elsevier}
}

@article{behrenfeld2019global,
  title={Global satellite-observed daily vertical migrations of ocean animals},
  author={Behrenfeld, Michael and Gaube, Peter and Della Penna, Alice and O’malley, Robert and Burt, William and Hu, Yongxiang and Bontempi, Paula and Steinberg, Deborah and Boss, Emmanuel and Siegel, David and Doney, Scott},
  journal={Nature},
  volume={576},
  number={7786},
  pages={257--261},
  year={2019},
  publisher={Nature Publishing Group UK London}
}

@article{grimes2014viewing,
  title={Viewing marine bacteria, their activity and response to environmental drivers from orbit: satellite remote sensing of bacteria},
  author={Grimes, D Jay and Ford, Tim E and Colwell, Rita R and Baker-Austin, Craig and Martinez-Urtaza, Jaime and Subramaniam, Ajit and Capone, Douglas G},
  journal={Microbial ecology},
  volume={67},
  number={3},
  pages={489--500},
  year={2014},
  publisher={Springer}
}

@article{legge2020carbon,
  title={Carbon on the northwest European shelf: Contemporary budget and future influences},
  author={Legge, Oliver and Johnson, Martin and Hicks, Natalie and Jickells, Tim and Diesing, Markus and Aldridge, John and Andrews, Julian and Artioli, Yuri and Bakker, Dorothee CE and Burrows, Michael T and others},
  journal={Frontiers in Marine Science},
  volume={7},
  pages={143},
  year={2020},
  publisher={Frontiers Media SA}
}

@article{telszewski2018biogeochemical,
  title={Biogeochemical in situ observations--motivation, status, and new frontiers},
  author={Telszewski, Maciej and Palacz, Artur and Fischer, Albert},
  journal={New Frontiers in Operational Oceanography},
  pages={131--160},
  year={2018}
}

@article{bakker2014update,
  title={An update to the Surface Ocean CO 2 Atlas (SOCAT version 2)},
  author={Bakker, Dorothee and Pfeil, Benjamin and Smith, Karl and Hankin, Steven and Olsen, Are and Alin, Simone and Cosca, Catherine and Harasawa, Sumiko and Kozyr, Alex and Nojiri, Yukihiro and Watson, A},
  journal={Earth System Science Data},
  volume={6},
  number={1},
  pages={69--90},
  year={2014},
  publisher={Copernicus Publications G{\"o}ttingen, Germany}
}

@article{racault2019environmental,
  title={Environmental reservoirs of Vibrio cholerae: challenges and opportunities for ocean-color remote sensing},
  author={Racault, Marie-Fanny and Abdulaziz, Anas and George, Grinson and Menon, Nandini and C, Jasmin and Punathil, Minu and McConville, Kristian and Loveday, Ben and Platt, Trevor and Sathyendranath, Shubha and Vijayan, V.},
  journal={Remote Sensing},
  volume={11},
  number={23},
  pages={2763},
  year={2019},
  publisher={MDPI}
}

@book{burchard1999gotm,
  title={GOTM, a general ocean turbulence model: Theory, implementation and test cases},
  author={Burchard, Hans and Bolding, Karsten and Villarreal, Manuel R},
  year={1999},
  publisher={Space Applications Institute}
}

@article{banerjee2025improved,
  title={Improved understanding of nitrate trends, eutrophication indicators, and risk areas using machine learning},
  author={Banerjee, Deep and Sk{\'a}kala, Jozef},
  journal={Biogeosciences},
  volume={22},
  number={15},
  pages={3769--3784},
  year={2025},
  publisher={Copernicus Publications G{\"o}ttingen, Germany}
}

@article{baretta1995european,
  title={The European regional seas ecosystem model, a complex marine ecosystem model},
  author={Baretta, JW and Ebenh{\"o}h, W and Ruardij, P},
  journal={Netherlands Journal of Sea Research},
  volume={33},
  number={3-4},
  pages={233--246},
  year={1995},
  publisher={Elsevier}
}

@article{bruggeman2014general,
  title={A general framework for aquatic biogeochemical models},
  author={Bruggeman, Jorn and Bolding, Karsten},
  journal={Environmental modelling \& software},
  volume={61},
  pages={249--265},
  year={2014},
  publisher={Elsevier}
}

@article{butenschon2016ersem,
  title={ERSEM 15.06: a generic model for marine biogeochemistry and the ecosystem dynamics of the lower trophic levels},
  author={Butensch{\"o}n, Momme and Clark, James and Aldridge, John N and Allen, Julian Icarus and Artioli, Yuri and Blackford, Jeremy and Bruggeman, Jorn and Cazenave, Pierre and Ciavatta, Stefano and Kay, Susan and Torres, Ricardo},
  journal={Geoscientific Model Development},
  volume={9},
  number={4},
  pages={1293--1339},
  year={2016},
  publisher={Copernicus GmbH}
}

@article{good2013en4,
  title={EN4: Quality controlled ocean temperature and salinity profiles and monthly objective analyses with uncertainty estimates},
  author={Good, Simon A and Martin, Matthew J and Rayner, Nick A},
  journal={Journal of Geophysical Research: Oceans},
  volume={118},
  number={12},
  pages={6704--6716},
  year={2013},
  publisher={Wiley Online Library}
}

@article{kay2019north,
  title={North West European Shelf Production Centre NORTHWESTSHELF\_ANALYSIS\_FORECAST\_BIO\_004\_011, Quality Information Document},
  author={Kay, Susan and McEwan, Robert and Ford, David},
  journal={Copernicus Marine Environment Monitoring Service},
  year={2019}
}

@article{king2018improving,
  title={Improving the initialisation of the Met Office operational shelf-seas model},
  author={King, Robert R and While, James and Martin, Matthew J and Lea, Daniel J and Lemieux-Dudon, Benedicte and Waters, Jennifer and O’Dea, Enda},
  journal={Ocean Modelling},
  volume={130},
  pages={1--14},
  year={2018},
  publisher={Elsevier}
}

@article{lenhart2010predicting,
  title={Predicting the consequences of nutrient reduction on the eutrophication status of the North Sea},
  author={Lenhart, Hermann-J and Mills, David K and Baretta-Bekker, Hanneke and Van Leeuwen, Sonja M and Van Der Molen, Johan and Baretta, Job W and Blaas, Meinte and Desmit, Xavier and K{\"u}hn, Wilfried and Lacroix, Genevi{\`e}ve and Wakelin, Sarah},
  journal={Journal of Marine Systems},
  volume={81},
  number={1-2},
  pages={148--170},
  year={2010},
  publisher={Elsevier}
}

@article{madec2015nemo,
  title={NEMO ocean engine},
  author={Madec, Gurvan and others},
  year={2015},
  publisher={Institut Pierre-Simon Laplace}
}

@article{mogensen2012nemovar,
  title={The NEMOVAR ocean data assimilation system as implemented in the ECMWF ocean analysis for System 4},
  author={Mogensen, Kristian and Balmaseda, Magdalena Alonso and Weaver, Anthony},
  year={2012},
  publisher={ECMWF Reading}
}

@article{o2017co5,
  title={The CO5 configuration of the 7 km Atlantic Margin Model: large-scale biases and sensitivity to forcing, physics options and vertical resolution},
  author={O'Dea, Enda and Furner, Rachel and Wakelin, Sarah and Siddorn, John and While, James and Sykes, Peter and King, Robert and Holt, Jason and Hewitt, Helene},
  journal={Geoscientific Model Development},
  volume={10},
  number={8},
  pages={2947},
  year={2017},
  publisher={Copernicus GmbH}
}

@article{sathyendranath2019ocean,
  title={An Ocean-Colour Time Series for Use in Climate Studies: The Experience of the Ocean-Colour Climate Change Initiative (OC-CCI)},
  author={Sathyendranath, Shubha and Brewin, Robert JW and Brockmann, Carsten and Brotas, Vanda and Calton, Ben and Chuprin, Andrei and Cipollini, Paolo and Couto, Andr{\'e} B and Dingle, James and Doerffer, Roland and Platt, Trevor},
  journal={Sensors},
  volume={19},
  number={19},
  pages={4285},
  year={2019},
  publisher={Multidisciplinary Digital Publishing Institute}
}

@article{siddorn2013analytical,
  title={An analytical stretching function that combines the best attributes of geopotential and terrain-following vertical coordinates},
  author={Siddorn, JR and Furner, R},
  journal={Ocean Modelling},
  volume={66},
  pages={1--13},
  year={2013},
  publisher={Elsevier}
}

@article{skakala2018assimilation,
  title={The assimilation of phytoplankton functional types for operational forecasting in the northwest European Shelf},
  author={Sk\'akala, Jozef and Ford, David and Brewin, Robert JW and McEwan, Robert and Kay, Susan and Taylor, Benjamin and de Mora, Lee and Ciavatta, Stefano},
  journal={Journal of Geophysical Research: Oceans},
  volume={123},
  number={8},
  pages={5230--5247},
  year={2018},
  publisher={Wiley Online Library}
}

@article{skakala2019improved,
  title={Improved representation of underwater light field and its impact on ecosystem dynamics: a study in the North Sea},
  author={Sk\'akala, Jozef and Bruggeman, Jorn and Brewin, Robert JW and Ford, David A and Ciavatta, Stefano},
  journal={Journal of Geophysical Research: Oceans},
  pages={e2020JC016122},
  year={2020},
  publisher={Wiley Online Library}
}

@article{skakala2020towards,
  title={Towards a multi-platform assimilative system for ocean biogeochemistry},
  author={Sk\'akala, Jozef and Ford, David Andrew and Bruggeman, Jorn and Hull, Tom and Kaiser, Jan and King, Robert R and Loveday, Benjamin Roger and Palmer, Matthew R and Smyth, Timothy James and Williams, Charlotte Anne June and Ciavatta, Stefano},
  journal={Journal of Geophysical Research: Oceans}, 
  volume={126(4)}, 
  pages={p.e2020JC016649.},
  year={2021},
  publisher={American Geophysical Union}
}

@article{skakala2022impact,
  title={The impact of ocean biogeochemistry on physics and its consequences for modelling shelf seas},
  author={Sk\'akala, Jozef and Bruggeman, Jorn and Ford, David and Wakelin, Sarah and Akp{\i}nar, An{\i}l and Hull, Tom and Kaiser, Jan and Loveday, Benjamin R and O’Dea, Enda and Williams, Charlotte AJ and others},
  journal={Ocean Modelling},
  volume={172},
  pages={101976},
  year={2022},
  publisher={Elsevier}
}

@article{skakala2024how,
  title={How uncertain and observable are marine ecosystem indicators in shelf seas?},
  author={Sk\'akala, Jozef and Ford, David and Fowler, Alison and Lea, Dan and Martin, Matthew J and Ciavatta, Stefano},
  journal={Progress in Oceanography},
  volume={224},
  pages={103249},
  year={2024},
  publisher={Elsevier}
}

@article{kay2016north,
  title={North West European Shelf Production Centre NWSHELF\_MULTIYEAR\_BIO\_004\_011},
  author={Kay, Susan and McEwan, Robert and Ford, David},
  journal={CMEMS Report},
  volume={3},
  pages={21},
  year={2021}
}

@dataset{kay2016north-data,
  author    = {Renshaw, Richard and Ford, David},
  title     = {Product Title. E.U. Copernicus Marine Service Information (CMEMS). Marine Data Store (MDS)},
  journal={[Dataset]},
  year      = {2025},
  doi       = {10.48670/moi-00058},
}

@article{cao2020diagnosis,
  title={Diagnosis of CO2 dynamics and fluxes in global coastal oceans},
  author={Cao, Zhimian and Yang, Wei and Zhao, Yangyang and Guo, Xianghui and Yin, Zhiqiang and Du, Chuanjun and Zhao, Huade and Dai, Minhan},
  journal={National Science Review},
  volume={7},
  number={4},
  pages={786--797},
  year={2020},
  publisher={Oxford University Press}
}

@article{chau2022seamless,
  title={A seamless ensemble-based reconstruction of surface ocean pCO 2 and air--sea CO 2 fluxes over the global coastal and open oceans},
  author={Chau, Thi Tuyet Trang and Gehlen, Marion and Chevallier, Fr{\'e}d{\'e}ric},
  journal={Biogeosciences},
  volume={19},
  number={4},
  pages={1087--1109},
  year={2022},
  publisher={Copernicus GmbH}
}

@article{dai2022carbon,
  title={Carbon fluxes in the coastal ocean: synthesis, boundary processes, and future trends},
  author={Dai, Minhan and Su, Jianzhong and Zhao, Yangyang and Hofmann, Eileen E and Cao, Zhimian and Cai, Wei-Jun and Gan, Jianping and Lacroix, Fabrice and Laruelle, Goulven G and Meng, Feifei and others},
  journal={Annual Review of Earth and Planetary Sciences},
  volume={50},
  number={1},
  pages={593--626},
  year={2022},
  publisher={Annual Reviews}
}

@article{roobaert2019spatiotemporal,
  title={The spatiotemporal dynamics of the sources and sinks of CO2 in the global coastal ocean},
  author={Roobaert, Aliz{\'e}e and Laruelle, Goulven G and Landsch{\"u}tzer, Peter and Gruber, Nicolas and Chou, Lei and Regnier, Pierre},
  journal={Global Biogeochemical Cycles},
  volume={33},
  number={12},
  pages={1693--1714},
  year={2019},
  publisher={Wiley Online Library}
}

@article{roobaert2024unraveling,
  title={Unraveling the physical and biological controls of the global coastal CO2 sink},
  author={Roobaert, Alizee and Resplandy, Laure and Laruelle, Goulven G and Liao, Enhui and Regnier, Pierre},
  journal={Global Biogeochemical Cycles},
  volume={38},
  number={3},
  pages={e2023GB007799},
  year={2024},
  publisher={Wiley Online Library}
}

%\appendix
\newpage

\section*{\Huge ~~~~Appendix A: Figures}

%\vspace{5cm}
\newpage

\begin{figure}
\hspace{-2cm}
\noindent\includegraphics[width=18cm, height=6cm]{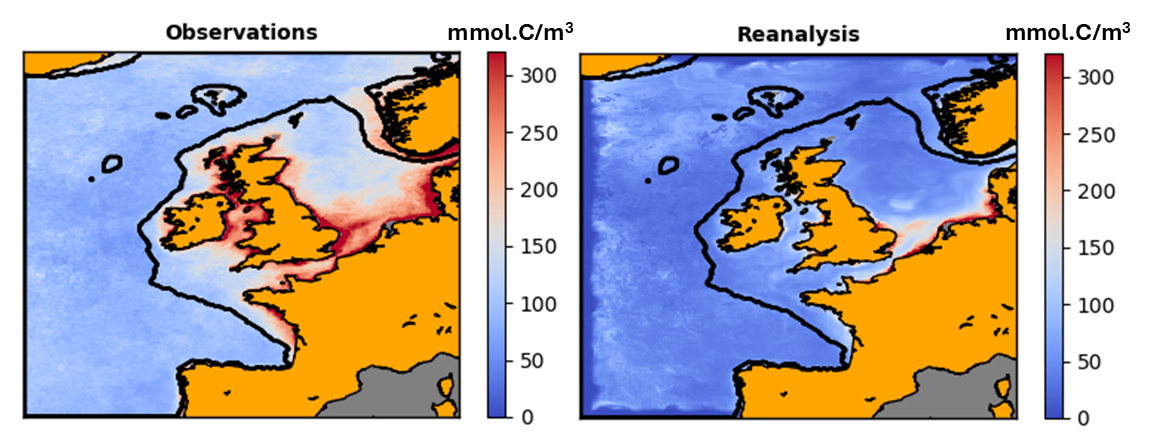}
\makeatletter 
\renewcommand{\thefigure}{A1}%\@arabic\c@figure}
\makeatother
\caption{Comparing the Ocean Color (OC) - Climate Change Initiative (CCI) v4.2 satellite-derived product for total particulate organic carbon concentrations (in mgC/m$^{3}$), broadly representing aggregate across phytoplankton, zooplankton, bacteria and detrital matter (left-hand panel), with the corresponding Copernicus reanalysis output (right-hand panel). The panels show temporally averaged values across the 2016-2017 period. The reanalysis was masked wherever there were missing satellite data, to ensure like-to-like comparison.}
\label{Fig.S1}
\end{figure}

\begin{figure}
\hspace{-2cm}
\noindent\includegraphics[width=16cm, height=10cm]{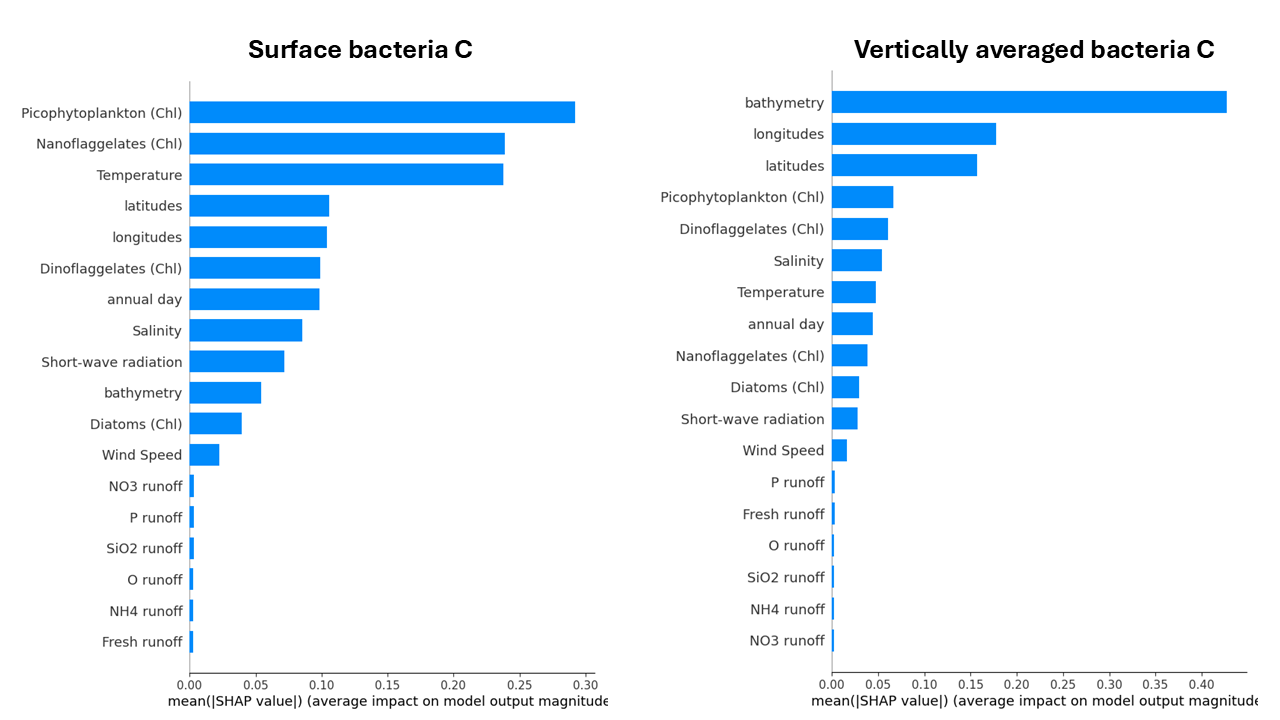}
\makeatletter 
\renewcommand{\thefigure}{A2}%\@arabic\c@figure}
\makeatother
\caption{SHapley Aditive exPlanations (SHAP) analysis (shown absolute values) for the surface heterotrophic bacteria carbon (left) and the vertically averaged heterotrophic bacteria carbon (right).}
\label{Fig.S2}
\end{figure}

\begin{figure}
\hspace{-2cm}
\noindent\includegraphics[width=16cm, height=10cm]{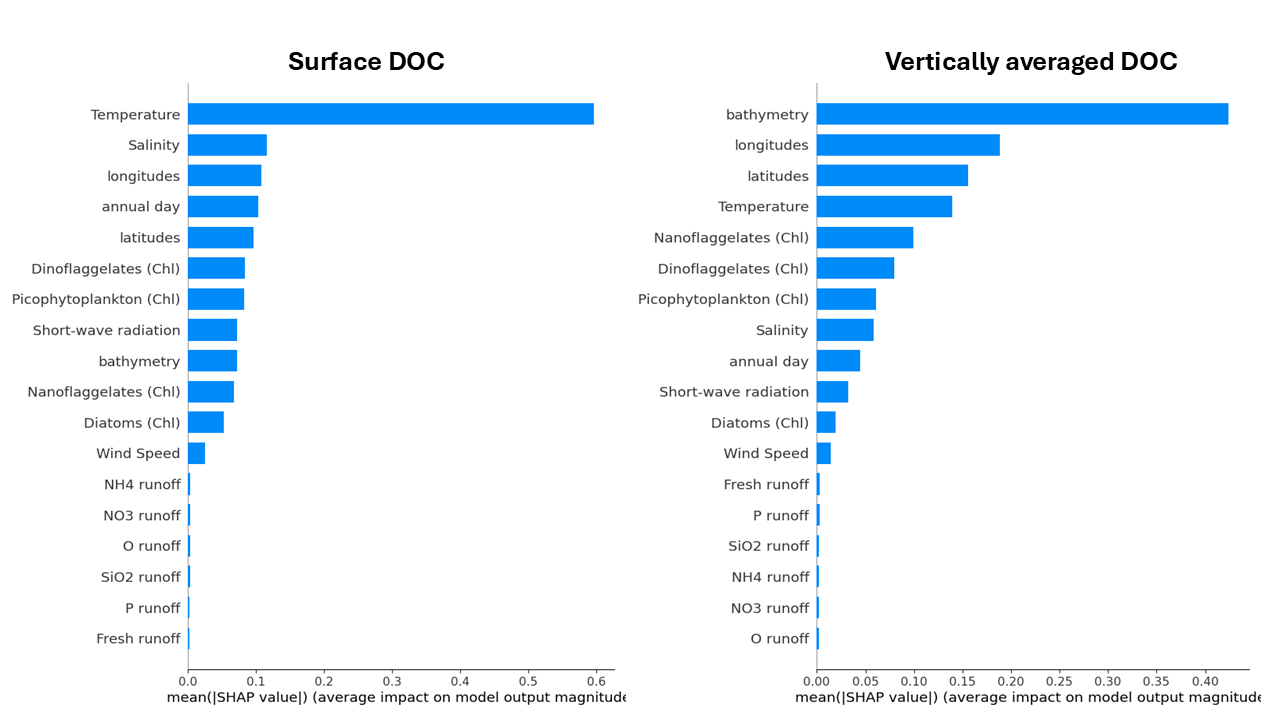}
\makeatletter 
\renewcommand{\thefigure}{A3}%\@arabic\c@figure}
\makeatother
\caption{SHAP analysis (shown absolute values) for the surface dissolved organic carbon (DOC) (left) and the vertically averaged DOC (right).}
\label{Fig.S3}
\end{figure}

\end{document}